\documentclass[aps,prb,nolongbibliography,notitlepage,twocolumn,superscriptaddress]{revtex4-2}

\usepackage{amsmath}
\usepackage{amssymb}
\usepackage{bm}
\usepackage{hyperref}

\usepackage[dvipsnames]{xcolor}
\usepackage{braket}
\usepackage{graphicx}
\usepackage{CJKutf8}
\def\Y{{\rm Y}}

\def\Q{{\rm Q}}

\def\P{{\rm P}}

\def\Q{{\rm Q}}

\begin{document}

\title{Anyon dispersion from non-uniform magnetic field on the sphere}
\author{Mina-Lou Schleith}
\thanks{These authors contributed equally}
\affiliation{Department of Physics, Harvard University, Cambridge, MA 02138, USA}
\affiliation{Institute for Theoretical Physics, ETH Zurich, CH-8093, Switzerland}
\author{Tomohiro Soejima (\begin{CJK*}{UTF8}{bsmi}副島智大\end{CJK*})}
\thanks{These authors contributed equally}
\affiliation{Department of Physics, Harvard University, Cambridge, MA 02138, USA}
\author{Eslam Khalaf}
\affiliation{Department of Physics, Harvard University, Cambridge, MA 02138, USA}
\date{\today}
\begin{abstract}
The discovery of fractional quantum anomalous Hall states in moir\'e systems has raised the interesting possibility of realizing phases of itenerant anyons. Anyon dispersion is only possible in the absence of continuous magnetic translation symmetry (CMTS). Motivated by this, we consider anyons on the sphere in the presence of a non-uniform magnetic field which breaks the ${\rm SU}(2)$ rotation symmetry, the analog of CMTS on the sphere, down to a ${\rm U}(1)$. This allows us to study the energy dispersion of the anyons as a function of $L_z$ angular-momentum, while maintaining the perfect flatness of single-particle dispersion. We parametrize the non-uniform field by a real parameter $R$ which concentrates the field at the north (south) pole for $R>1$ ($R < 1$), and show that, for our choice of field, any $p$-body correlation function evaluated in the space of Laughlin quasiholes can be mapped \emph{exactly} to a corresponding $p$-body correlation function in uniform field. In the thermodynamic limit, this enables us to analytically compute the interaction-generated spatially varying potential felt by the anyons. Remarkably, such spatially varying potential is sufficient to generate dispersion for the anyons, which we compute exactly, up to an overall scaling constant. The anyon dispersion in our model describes azimuthal motion around the sphere at a constant height, similar to spin precession. Our work therefore serves as a concrete demonstration that interaction alone can generate nonzero anyon dispersion in the presence of inhomogeneous magnetic field.
\end{abstract}
\maketitle

\emph{Introduction}--- The fractional quantum anomalous Hall effect (FQAHE), a zero-field version of the fractional quantum Hall (FQH) effect, was recently observed in two distinct material platforms: twisted MoTe$_2$~\cite{Cai_2023, Park2023, Xu_2023, Zeng_2023} and rhombohedral pentalayer graphene on hexagonal boron nitride~\cite{Lu2024}. From a practical viewpoint, the realization of the FQH effects at zero magnetic field has been a longstanding goal for material design due its potential for anyon-based quantum computation~\cite{kitaev2003fault}
without the need for impractically large magnetic fields. However, besides their practical relevance, one natural question is whether they can harbor qualitatively new physics beyond what is possible in Landau levels (LLs).

The physics of LLs is strongly constrained by their symmetry; continuous magnetic translation symmetry (CMTS) acts on both position and momentum, ensuring uniform real-space charge distribution and momentum-space quantum geometry (Berry curvature, quantum metric, etc). Thus, CMTS prohibits finite dispersion for any charged excitation. This implies not only vanishing single-particle dispersion, but also vanishing dispersion for many-body charged quasiparticles such as anyons. In contrast to LLs, fractional Chern insulators (FCIs) defined on the lattice~\cite{liuRecentDevelopmentsFractional2022,parameswaranFractionalQuantumHall2013,BergholtzReview2013,neupertFractionalQuantumHall2011,shengFractionalQuantumHall2011,regnaultFractionalChernInsulator2011,qi_generic_2011,parameswaranFractionalChernInsulators2012,wuBlochModelWave2013,kourtisFractionalChernInsulators2014,YahuiChern,tarnopolskyOriginMagicAngles2019,ledwithFractionalChernInsulator2020a,CecilleFCI,AbouelkomsanTBG,AbouelkomsanTDBG,meraEngineeringGeometricallyFlat2021,Wang_2021,ledwithFamilyIdealChern2022,vortexability}, including FQAHE, have only discrete translation symmetry, thereby allowing nonzero dispersion \cite{shi2024, RegnaultDispersion}. Anyon dispersion opens the door to the realization of exotic itinerant anyon phases upon doping FCI states, including anyon superconductors  \cite{Laughlin_1988, Fetter_1989, Lee_1989, Chen_1989, Halperin_1989, Wen_1990, Wen_1991,Tang_2013,kim2025topological, shi2024,divic_2024, shi2025dopinglatticenonabelianquantum, shi2025anyon, nosov2025anyonsuperconductivityplateautransitions, ClemensAnyonSC}. This leads naturally to the question: what is the microscopic origin of anyon dispersion?

A straightforward way to break CMTS is to introduce a periodic potential, but this has several drawbacks. First, the potential-induced kinetic energy of electrons competes with interaction, potentially destabilizing the formation of FCI states \cite{kousa2025theory, reddy2023fractional, paul2025shining}. This means it is difficult to form a sizable dispersion without destroying the parent state. Second, the potential prevents us from writing exact ground states, hindering analytical progress.

An alternative approach is to consider a non-uniform magnetic field, which can break CMTS without inducing single-particle dispersion. In fact, recent works have shown that the flat Chern bands emerging in moir\'e systems can, under certain idealized assumptions, be mapped exactly to the lowest Landau level (LLL) of a Dirac particle in a non-uniform magnetic field \cite{tarnopolskyOriginMagicAngles2019, ledwith2020fractional, Wang_2021, morales-duran2024magic, TMDAharonocCasher, dong2023composite}. Such bands, dubbed ideal or Aharonov-Casher bands \cite{tarnopolskyOriginMagicAngles2019,ledwithFractionalChernInsulator2020a,meraEngineeringGeometricallyFlat2021,meraKahlerGeometryChern2021,Wang_2021,ledwith2022family, ledwith_vortexability_2023, aharonov1979ground, TMDAharonocCasher}, share the same holomorphic structure of the LLL, which ensures the realization of FCI states for sufficiently short-range interactions \cite{ledwith2020fractional, Wang_2021, ledwith2023vortexabilitya}. On the other hand, they feature non-uniform real-space charge distribution and momentum-space quantum geometry due to the absence of CMTS. 

Non-uniform quantum geometry can induce anyon dispersion in the following way. Consider the wavefunction of anyon localized at location $\xi$. Since we are setting the kinetic energy to zero, the energy of this anyon is determined by interactions alone. Non-uniformity of quantum geometry implies the effective interaction felt by the anyon is spatially dependent, resulting in an effective potential $V[\xi]$ for anyons. Since anyons feel a background magnetic field, the two components of the position operator do not commute, i.e. the $y$-coordinate acts as the momentum conjugate to $x$. As a result, a pure potential energy term generates a kinetic energy, leading to anyon dispersion.

In this work, we provide an explicit demonstration of this mechanism by studying anyon dispersion \emph{on the sphere} in the presence of a non-uniform magnetic field. One of the main advantages of spherical geometry is its simplicity~\cite{Haldane_1983} which will enable us to compute anyon dispersion analytically under some assumptions. To understand how the dispersion on the sphere is related to the standard band case (torus), note that the analog of CMTS on the sphere is SU(2) rotation symmetry. On the torus, CMTS is broken by introducing a periodic magnetic field whose unit cell contains $2\pi$ flux. This ensures momentum remains a good quantum number. Analogously, we break SU(2) on the sphere by introducing a non-uniform magnetic field which only depends on the azimuthal angle, leaving an unbroken U(1) symmetry generated by $L_z$. This enables us to label states using $L_z$ eigenvalues.  

We numerically show that the anyon energies depend on angular momentum, corresponding to anyons that move around the $z$ axis. Remarkably, we find a choice for non-uniform magnetic field which enables us to compute the anyon dispersion from the knowledge of two-body functions \textit{in the uniform field}. This allows us to develop an analytic understanding of the anyon energetics, leading to an analytic formula for the anyon dispersion in the thermodynamic limit. Our work therefore serves as a concrete demonstration that interaction alone can generate nonzero anyon dispersion in the presence of non-uniform magnetic field or quantum geometry. We anticipate that, while the physics of our setup is likely distinct from that of periodically modulated magnetic field, relevant to ideal flat bands, the energetic origin for the dispersion in the two setups is the same.

\emph{LLL on the sphere with inhomogeneous magnetic field}--- We consider LLL on the unit sphere with a charge $2s$ Dirac monopole placed at its center \cite{Haldane_1983}. The resulting radial magnetic field is $\bm{B} = B(\theta, \phi) \hat{\bm\Omega}$, where $\hat{\bm\Omega}$ is a radial unit vector, and $\Phi_0 = h/e$ is the flux quantum. We will interchangeably use the standard spherical coordinates $(\theta, \phi)$, or the stereographically projected coordinates $z=\cot(\theta/2)e^{i\phi}$. The latter make the holomorphic structure of the wavefunctions more transparent.
 The uniform case is given by $\bm{B} = 2s\Phi_0 \hat{\bm\Omega}/4\pi$.
A non-uniform magnetic field can be characterized by its associated magnetic potential $Q^R(z)$, where $R>0$ is a parameter that controls the inhomogeneity. The magnetic potential in the planar coordinates satisfies $\partial_z \partial_{\bar z} Q^R(|z|) =
-\frac{\pi \sqrt{g}}{\Phi_0} B(|z|)$, where $g$ is the metric.
The associated Hamiltonian takes the form
\begin{equation}
    H=-\frac{2}{\sqrt{g}} \biggl( \partial_z - \frac{1}{2}\partial_z Q^R(|z|) \biggr) \biggl( \partial_{\bar z} + \frac{1}{2} \partial_{\bar z} Q^R(|z|) \biggr) + \frac{2-g_s}{4} B,
\end{equation}
where $\sqrt{g} = 4/(1+|z|^2)^2$. Setting $g_s=2$ yields the Aharonov-Casher Hamiltonian whose zero modes are \cite{fDirac}
\begin{equation}
     \phi^R_{s,m}(z,\bar z) = e^{-\frac{\beta_m^R}{2}}\alpha_{s,m} z^{s+m} e^{\Q^R(|z|)/2}. \label{Haldane_spinor_wave functions}
\end{equation}
where $m = -s,-s+1,\dots,s$, $\alpha_{s, m} =  \sqrt{\frac{(2s+1)!}{4\pi\, (s-m)!(s+m)!}}$, and $\beta^R_m$ is a normalization constant. When $Q^R$ is symmetric about the rotation around the $z$ axis, $m$ coincides with the $L_z$ angular momentum quantum number. 

\begin{figure}
    \centering
    \includegraphics[width=0.9\linewidth]{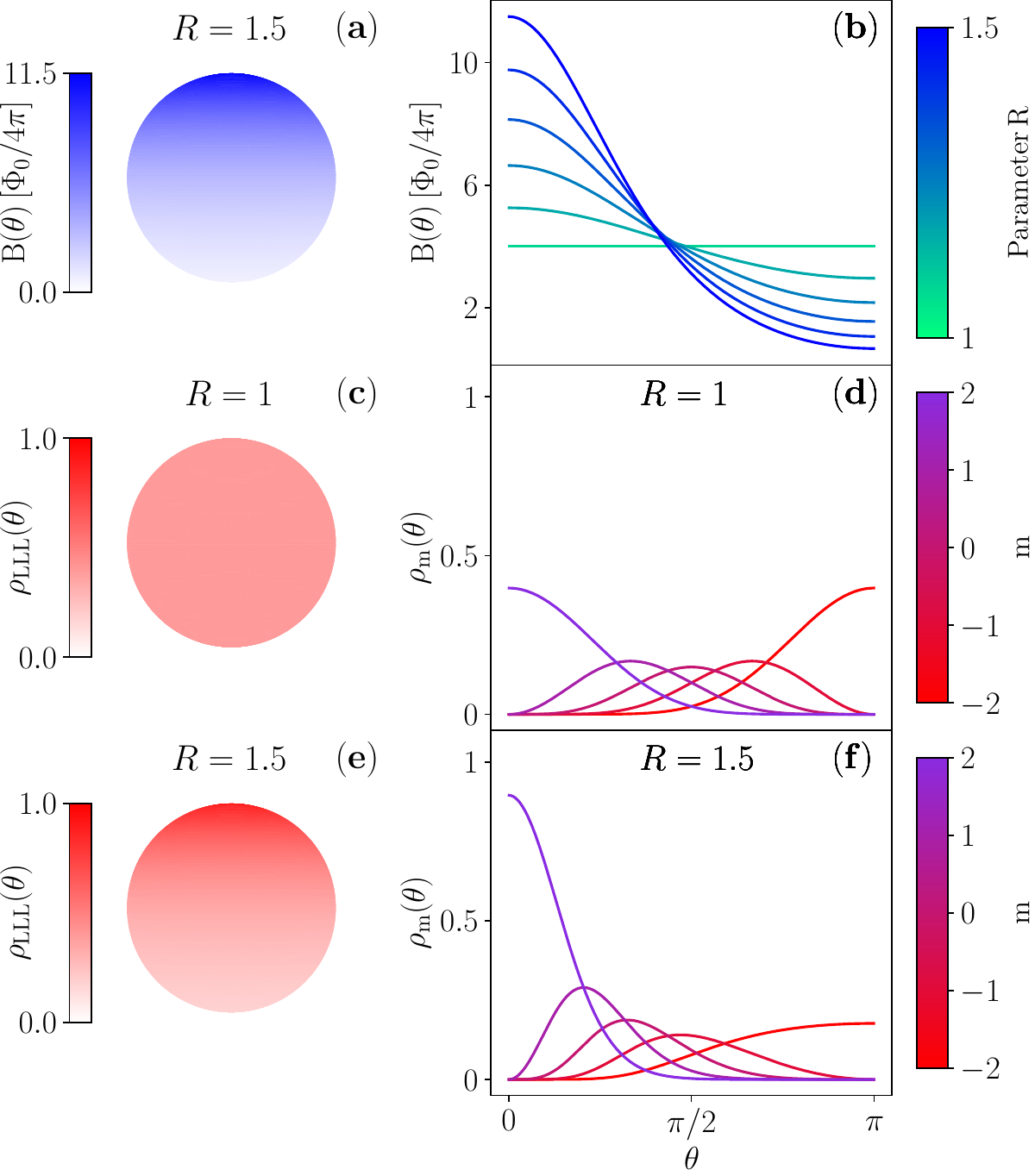}
    \caption{\textbf{Non-uniform magnetic flux density and LLL charge distributions} on the spherical geometry with total flux $\Phi=2s\Phi_0=4\Phi_0$. (a,b) The magnetic flux density for $R=1.5$ and $R \in [1,1.5]$ respectively. (c-f) The total charge density $\rho_{\rm LLL}(\theta)$ as well as the charge density $\rho_m(\theta)$ of different $m=-2,\dots,2$ eigenstates in the LLL for $R=1$ and $R=1.5$. }
    \label{fig:magfield}
\end{figure}

We consider the following rotationally symmetric magnetic potential parameterized by $R>0$:
\begin{equation}
     Q^R(|z|) = 2 \log (1 + |z|^2) - 2 (s + 1) \log(1 + |z|^2/R^2)
     \label{QR}.
 \end{equation}
 Its associated magnetic field is given by
 \begin{equation}
    B^R(|z|) = \frac{\Phi_0}{4\pi} \biggl( 2(s+1) \frac{(1+|z|^2)^2}{(1+|z|^2/R^2)^2} - 2 \biggr),
    \label{NonUniformField}
\end{equation} 
which reduces to the uniform case at $R=1$. For $R > 1$ ($R < 1$), this field is more concentrated at the north (south) pole as shown in Fig.~\ref{fig:magfield} (a,b). The two cases are related by the inversion $z \mapsto 1/z$, so for the remainder of this paper, we will restrict ourselves to $R \geq 1$.
 With this choice of $Q^R$, we find the normalization constant $\beta_m^R = 2(s+m+1) \log R$. 
Note that we chose $\alpha_{s, m}$ such that $\beta_m^{R=1} = 0$. The effect of $R > 1$ on the single-particle wave functions is illustrated in Fig.~\ref{fig:magfield} (c-f). The states with $m > 0$ are near the north pole and get more squeezed whereas the states with $m < 0$ are near the south pole and become more widely spread.
 
To setup the interacting problem, we consider the LLL-projected spherically symmetric density-density interaction
\begin{equation}
    \hat V = \int d\mu(z_1) d\mu(z_2) V(d(z_1, z_2)) :\hat \rho^R_{z_1} \hat \rho^R_{z_2}:,
\end{equation}
where $d\mu(z_i) = 4 dx_i dy_i / (1 + |z_i|^2)^2$ is the integration measure, and $\hat \rho^R_z = \hat c_z^{\dagger, R} \hat c^R_z$ is the LLL-projected density operator with $\hat c^R_z = \sum_{m=-s}^s \phi^R_{s,m}(z,\bar z) \hat c_m$ and $ \hat c_m$ is the annihilation operator for a state with angular momentum $m$. To enforce spherical symmetry, the interaction depends only on the distance function $d(z_1, z_2)$ defined as $
    d(z_1, z_2) = \frac{2|z_1 - z_2|}{\sqrt{1+|z_1|^2}\sqrt{1+|z_2|^2}}.
$ which is the chord distance function on the sphere $2\sin(\Omega_{12}/2)$, where $\Omega_{12}$ is the relative angle between the two points.

As representative interactions, we consider Trugman-Kivelson pseudopotentials~\cite{trugman-kivelson} $V^n(z_1, z_2) = \frac{1}{(n!)^2}\Delta_g^{n} \delta_g(z_1, z_2)$, where $\Delta_g = \frac{4}{\sqrt{g(z)}}  \partial_z \partial_{\bar{z}}$ and $\delta_g(z,z_0) = \delta(d(z, z_0)) $, as well as the Coulomb interaction $V^{\mathrm{Coulomb}}
(z_1, z_2) = 1/d(z_1, z_2)$.
A derivation of the second quantized form of these Hamiltonians is detailed in App.~B.

\begin{figure}
    \centering
    \includegraphics[width=\linewidth]{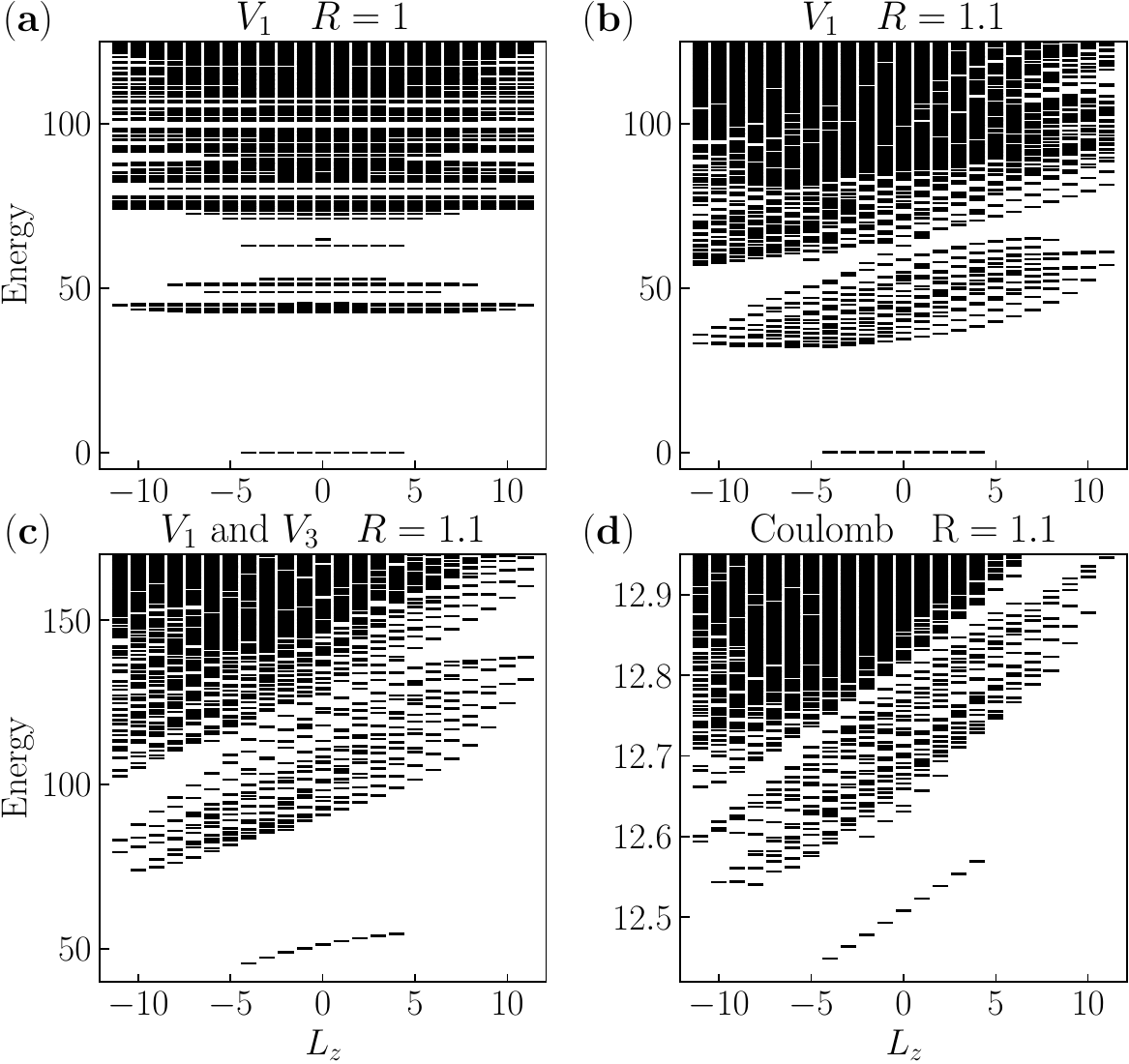}
    \caption{\textbf{ED spectra of anyons in the $\nu=\frac{1}{3}$ LLL on the sphere} for $N=8$ electrons at total flux $\Phi=2s\Phi_0=22\Phi_0$ (Laughlin $\frac{1}{3}$ plus one quasihole). (a,b) $V_1$ pseudopotential spectrum at $R=1$ and $1.5$ showing a $9$-fold degenerate quasihole manifold. (c,d)  $V_1+ (3!)^2 \times 10^{-4} \times V_3$ and Coulomb pseudopotential interactions lift the degeneracy, producing a dispersive quasihole band.}
    \label{fig:AnyonDisp}
\end{figure}

 \emph{Exact diagonalization}--- To establish the dependence of the anyon energies on their angular momentum for $R \neq 1$, we perform finite-size ED on the spherical geometry with electrons in the LLL. We consider magnetic flux of $2s\Phi_0 = (3(N-1) + 1)\Phi_0$, where $N$ is the number of electrons, corresponding to one extra flux quantum on top of $\nu = 1/3$ leading to the creation of one quasihole. In Figure~\ref{fig:AnyonDisp}, we show the ED spectra for $N = 8$ ($2s = 22$) for different interactions.
 
 If we only include the first pseudopotential $V_1$, we find $N+1=9$ zero modes labeled by the total angular momentum $L_z$ (panel (a)) that can be associated with  individual root configurations of the quasi-hole~\cite{Bernevig_2008, Yang_2012}. These remain zero modes for non-uniform field (panel (b)). Once we include interactions beyond $V_1$ we find that this set of zero modes acquire a finite energy that becomes dispersive for non-uniform field. For small $V_3$ interaction in panel (c), the dispersion is concave, with larger slope at smaller angular momentum, whereas the slope is approximately linear for Coulomb in panel (d). 
 
\emph{Mapping to uniform field}--- 
We now show that the dispersion of  the quasiholes with non-uniform field can be understood from correlation functions of quasi-holes at $R=1$. We proceed by assuming the interaction is short-ranged such that we can write $\hat V = \hat V_1 + \Delta \hat V$ with sufficiently small $|\Delta \hat V|/|\hat V_1|$.
The dispersion in this limit can be obtained by the first order perturbation theory: $E_m = \bra{\Psi_m} \Delta V \ket{\Psi_m} / \braket{\Psi_m | \Psi_m}$, where $\ket{\Psi_m}$ is the ground state for $V_1$ with angular momentum $m$.

The zero modes of $V_1$ can be constructed exactly along the usual line of argument for the pseudopotential interaction~\cite{trugman-kivelson}.
The first pseudopotential $\hat V_1 = \nabla_g^2 \delta_g(z_1, z_2)$ penalizes any pair of particles with relative angular momentum 1. This leads to $N+1$ distinct zero modes with angular momenta $M = -\frac{N}{2},\dots,\frac{N}{2}$ whose wavefunctions are given by:
\begin{equation}
    \Psi^R_{M}(\{z_i\}) = \P_{M+N/2}(\{z_i\}) \prod_{i<j} (z_i - z_j)^3 \prod_i e^{\frac{1}{2} \Q^R(|z_i|)}.
    \label{eq:laughlin}
\end{equation}
Here, $\P_k(\{z_i\})$ is an $k^{\rm th}$ order symmetric polynomial obtained by expanding $\prod_{i=1}^N (z_i - w) = \sum_{k=0}^N P_k(\{z_i\}) (-w)^{N-k}$. 
Clearly, we can relate the wave functions at arbitrary $R$ to those at $R = 1$ via $\Psi_M^R(\{z_i\}) = \Psi_M(\{z_i\}) \prod_i \Gamma_R(z_i)$ where $\Gamma_R(z_i) = e^{\frac{1}{2} (\Q_R(|z_i|) - \Q(|z_i|))}$ and dropping the $R$ index indicates the uniform case $R = 1$.

The second quantized form of these wavefunctions are given by
\begin{equation}
    \ket{\Psi^R_M} = \int \prod_i d\mu(z_i) \Psi^R_{M}(\{z_i\}) \prod_i c_{R;z_i}^\dagger \ket{0}
\end{equation}
where $c_{R;z}^\dagger = \sum_{m} \phi^R_{m}(z)c^{\dagger}_{m}$.
We show in App.~C that the second quantized wavefunctions are related to each other by a simple relation  $|\Psi_M^R \rangle = \hat T_R |\Psi_M \rangle$ where $ \hat T_R$ is the invertible but non-unitary transformation $\hat T_R = \hat T_R^\dagger = e^{\frac{1}{2}\sum_m\beta^R_m  \hat c_m^\dagger \hat c_m}$. These relations do not depend on the form of the non-uniform field.

Let us now rewrite the energy for a quasihole with angular momentum $M$ as \footnote{Since we are working in the space of zero modes of $\hat V_1$, we can use $\hat V$ and $\Delta \hat V$ interchangeably}
\begin{equation}
    \epsilon_M^R = \int d\mu(z_1) d\mu(z_2) V(z_1,z_2) g_M^R(z_1,z_2)
\end{equation}
where $g_M^R(z_1,z_2):= \frac{\langle \Psi_M^R|c_{R;z_1}^\dagger c_{R;z_2}^\dagger c_{R;z_2} c_{R;z_1}|\Psi_M^R \rangle}{\langle \Psi_M^R|\Psi_M^R \rangle}$ is the pair correlation function. Using $c_{R;z} = \Gamma(z) \hat{T}_R c_z \hat{T}_R^{-1}$, we can write this as
\begin{equation}
    g_M^R(z_1,z_2) = \Gamma_R(z_1)^2 \Gamma_R(z_2)^2 \frac{\langle \Psi_M|c_{z_1}^\dagger c_{z_2}^\dagger \hat T_R^2 c_{z_2} c_{z_1}|\Psi_M \rangle}{\langle \Psi_M| \hat{T}_R^2 | \Psi_M \rangle}
\end{equation}
In general, this requires evaluating a $N$-body operator, since $\hat{T}_R$ acts on all electrons at once.
For the non-uniform field we chose (Eq.~\ref{NonUniformField}), however, we have  $\hat T_R = R^{(s+1) \hat N + \hat L_z}$. Therefore, the action of $\hat{T}_R$ can be reduced to a multiplication by a function of conserved quantum numbers. This reduces the evaluation of the numerator to that of a collection of two-body operator by expanding $c_z$ in angular momentum basis. This simplification generalizes to any $m$-body operator \cite{fWang}.

\emph{Transforming the interaction}--- We now derive explicit formulae for $\epsilon_M^R$. We first note that, by explicit computation (see App.~B), we have
\begin{align}
    \hat V^R &=\int d\mu(z_1) d\mu(z_2) V(d(z_1, z_2)) c^\dagger_{R;z_1} c^\dagger_{R;z_2} c_{R;z_2} c_{R;z_1}, \nonumber
    \\
    =   &\int d\mu(z_1) d\mu(z_2) V(d(Rz_1, Rz_2)) c^\dagger_{z_1} c^\dagger_{z_2} c_{z_2} c_{z_1},
    \label{eq:R_transformation}
\end{align}
with associated change in the distance function $
        d(R z_1, R z_2) = R \sqrt{h_R(z_1) h_R(z_2)} d(z_1, z_2)$
where $h_R(z) = (1 + |z|^2)/(1 + R^2 |z|^2)$. Thus, the role of the non-uniform field can be \emph{exactly} absorbed into non-uniform interaction through a dilation $z \mapsto R z$.
Since this is not a unitary transformation, the spectrum of the new Hamiltonian will be generally different from the old Hamiltonian. The $n$-th pseudopotential transforms as
\begin{equation}
    V^n(z_1, z_2) \to (R \sqrt{h_R(z_1) h_R(z_2)})^{-(2n+2)} V^n(z_1, z_2).
\end{equation}
Therefore, the transformed pseudopotential shares the same set of zero modes as the original pseudopotential.
We thus simply need to evaluate the energy of the zero modes with respect to this modified Hamiltonian, leading to
\begin{equation}
    \epsilon_M^R = \int d\mu(z_1) d\mu(z_2) V(R z_1,R z_2)  g_M(z_1,z_2),
\end{equation}
where $g_M(z_1, z_2)$ is the pair correlation function at $R=1$.

We can understand the role of non-uniform field from Eq.~\eqref{eq:R_transformation} as follows: The distances between particles is squeezed by $1/R$ for large $z$ ($\theta \approx 0$, north pole) and expanded it by $R$ for small $z$ ($\theta \approx \pi$, south pole). For any interaction that decays with distance, this makes the interactions stronger close to the north pole. As a result, quasiholes with large angular momentum whose wavefunctions have more weight on the north pole pay more energy, consistent with numerics.

The above mapping already allows us to relate $g_M^{R=1}(z_1, z_2)$ with $\epsilon_M^R$. It can be further simplified by considering the wavefunctions for a quasi-hole localized at point $\xi$:
\begin{equation}
    \Psi_\xi(z_1,\dots,z_N) = C_\xi \prod_i (z_i - \xi) \prod_{i<j} (z_i - z_j)^3 \prod_i e^{\frac{1}{2} Q^R(|z_i|)}
    \label{eq:laughlin_quasihole},
\end{equation}
where $\xi$ is the position of the quasihole, and $C_\xi:= (1 + |\xi|^2)^{-N/2}$ is the normalization constant. The wavefunctions $|\Psi_\xi \rangle$ can be understood as a spin coherent state defined as $|\Psi_\xi \rangle = \frac{\xi^{N}}{(1 + |\xi|^2)^{N/2}} e^{-\xi^{-1} L_+} |\Psi_{-N/2} \rangle$, which can be written as a linear superposition of angular momentum eigenstates $|\Psi_M \rangle = \sqrt{\frac{(N/2-M)!}{N!(N/2+M)!}} L_+^{M+N/2} |\Psi_{-N/2} \rangle$ by expanding the exponential. We can define the effective potential felt by quasihole at $\xi$ via
\begin{equation}
    v_\xi^R = \int d\mu(z_1) d\mu(z_2) V(R z_1,R z_2)  g_\xi(z_1,z_2).
    \label{eq:Potential}
\end{equation}
Using the relation between $|\Psi_\xi \rangle$ and $|\Psi_M \rangle$, we can extract the dispersion $\epsilon_M^R$ from $v_\xi^R$ by inverting the relation
\begin{equation}
\label{vtoe}
    v_\xi^R = \frac{1}{(1 + |\xi|^2)^{2l}} \sum_{m=-l}^l \binom{2l}{l-m} |\xi|^{2(l-m)} \epsilon_m^R.
\end{equation}
In the thermodynamic limit, $N \rightarrow \infty$, we can define a continuum version of angular momentum $\kappa = 2M/N \in [-1,1]$ for which we can derive a much simpler inversion formula
\begin{equation}
    \epsilon_\kappa^R = v^R_{\xi = \sqrt{\frac{1 - \kappa}{1 + \kappa}}}.
    \label{SimpleInversionFormula}
\end{equation}
This can be further simplified by noting that different quasihole locations are related to each other by a rotation of the sphere, which means that we only need to know $g_\xi(z_1, z_2)$ for a single fixed $\xi$ e.g. $\xi = 0$ corresponding to the south pole (see App.~D for details). The quasihole pair correlation function $g_{\rm SP}$ can be obtained numerically for large system sizes by performing Monte Carlo~\cite{girvin1984anomalous, fulsebakke2023parametrization}.

\emph{Coherent state path integral representation}--- One way to understand the origin of the dispersion and its relation to the potential $v_\xi$ is to use the observation that the state $|\Psi_\xi \rangle$ is a coherent state under the action of SU(2) and derive a coherent state path integral. The resulting Lagrangian takes the form $\mathcal L = \mathcal L_B - v_\xi$, where $\mathcal L_B$ is given by 
\begin{equation}
    \mathcal L_B = i \langle \Psi_\xi|\partial_t \Psi_\xi \rangle = -i \frac{N}{2} \frac{\bar \xi \dot \xi - \xi \bar {\dot \xi}}{1 + |\xi|^2} = \frac{N}{2}(1 - \cos \theta) \dot \phi,
\end{equation}
which is the Berry phase term for a spin $\frac{N}{2}$. Thus, we can interpret this Lagrangian as the Lagrangian of a spin $\frac{N}{2}$ particle subject to a potential that only depends on $S_z$, an analog of the spin precession problem. We can also interpret this Berry phase as that of a particle moving in the field of a magnetic monopole with charge $N$. This reflects the fact that the anyon sees a reduced magnetic flux that is roughly 1/3 of the one seen by the electrons. Quantizing this theory leads to a Hamiltonian, which is only a function of $S_z$, whose spectrum is related to the potential via (\ref{vtoe}). The simplified expression at large $N$, Eq.~\ref{SimpleInversionFormula}, is the corresponding semiclassical limit where we simply replace $\cos \theta$ in $v_\xi$ by $S_z$.

\begin{figure}
    \centering
     \includegraphics[width=\linewidth]{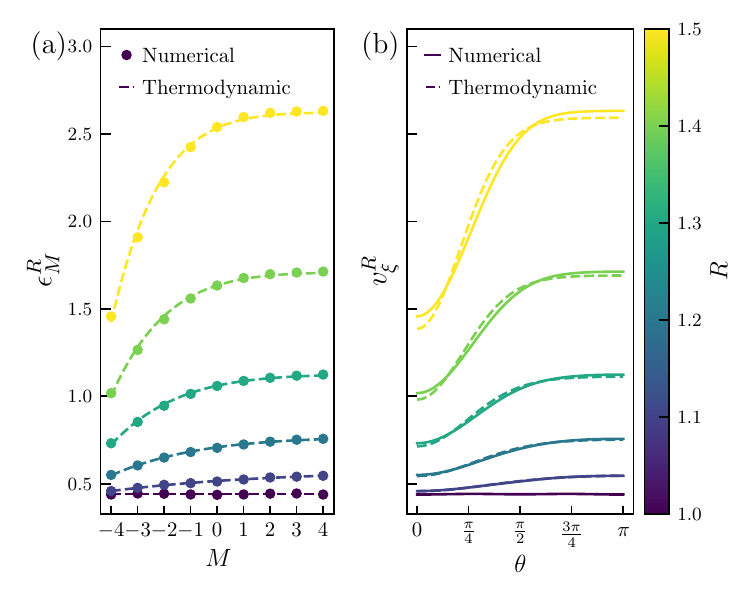}
    \caption{\textbf{Comparison of analytical and numerical anyons dispersion for short-range interaction} Numerical results were computed with $N=8$ ED with the interaction $V_1 + (3!)^2 \times 10^{-6} V_3$. Analytic results in the thermodynamic limit were obtained by fitting the numerical data using Eq.~\eqref{eq:qh_xi_analytic}, Eq.~\eqref{eq:qh_xi_analytic}.
    (a) Numerical and analytical results for $\epsilon_M^R$. (b) Numerical and analytical results for $v_\xi^R$ in terms of $\theta$.}
    \label{fig:dispersion_comparison}
\end{figure}

\emph{Thermodynamic limit}--- Rather remarkably, we can derive the asymptotic form of the dispersion in the thermodynamic limit by noting that the pair correlation function depends on $z_1, z_2$ via three distance functions $d(z_1, z_2)$, $d(z_1, \xi)$ and $d(z_2, \xi)$ and that its dependence saturates at distances large compared to magnetic length $\ell_b$ due to plasma screening. As shown in App.~E, this enables us to evaluate the potential energy for a quasihole at $\xi$:
\begin{equation}
v_\xi^R = a_0(R) - a_3 \left(\frac{1 + R^2 |\xi|^2}{R(1 + |\xi|^2)}\right)^8,
\label{eq:qh_xi_analytic}
\end{equation}
where $a_0(R)$ is an $R$-dependent offset and $a_3$ is independent of $\xi$ and $R$ \footnote{We choose the negative sign since we anticipate the function $\delta g_\xi(z_1, z_2)$ to be mostly negative as it corresponds to a dip relative to the asymptotic value}. 
From this, we can extract the full spectrum as
\begin{equation}
\epsilon^R_{\kappa} = a_0(R) - a_3 \left(\frac{1+\kappa + R^2(1-\kappa)}{2R}\right)^8
\label{eq:qh_m_analytic}
\end{equation}
where 
 $\kappa = 2m/N \in [-1, 1]$ as before. The parameter $a_3$ which controls the dispersion can be determined by fitting the bandwidth. The results for the fit of $\epsilon_M^R$ and $v_\xi^R$ for different values of $R$ are shown in Fig.~\ref{fig:dispersion_comparison}, showing excellent agreement with the numerics. For long-range Coulomb interaction, we can perform a similar analysis to obtain a similar, but more complicated expression (see App.~E).

\emph{Discussion}--- 
Our exact mapping reveals that inhomogeneous magnetic field effectively serves as a spatially varying potential for the quasihole. Thus, our setup resembles the physics of an FQHE droplet in a confining potential~\cite{xie1993elementary} and our states are the anyonic analog of Fock-Darwin states. However, we emphasize that our dispersion is generated purely from interactions, while the single-particle dispersion remains flat. It is worth noting that the dynamics of anyons in our model are distinct from those in a periodic potential. For instance, a quasihole localized at a point on the sphere can only move in the azimuthal direction similar to spin precession and unlike true two-dimensional dispersion. This suggests that the physical consequences of dispersion are likely different and the model may not harbor phases such as anyon superconductor. On the other hand, we expect the energetic mechanism for the generation of an effective potential that leads to anyon dispersion to be generic and applicable to the case of non-uniform magnetic field on the torus, which is relevant to flat bands with non-uniform quantum geometry. We leave a detailed analysis to future works. 

Our model allows us to extract the dependence of any correlation function of quasiholes on the non-uniform magnetic field by computing the correlation functions in the uniform field. This can enable a systematic study of how interaction between anyons is modified in the presence of non-uniform magnetic field and allows us to tackle questions such as the formation of multi-anyon bound state \cite{xu2025dynamicsclustersanyonsfractional, gattu2025molecularanyonsfractionalquantum} for non-uniform field.

\begin{acknowledgements}
We thank fruitful discussions with Junkai Dong, Zhaoyu Han, and Pavel A. Nosov. E. K. is supported by NSF MRSEC DMR2308817 through the Center for Dynamics and Control
of Materials. This research is funded in part by the Gordon and Betty Moore Foundation’s EPiQS Initiative,
Grant GBMF8683 to T.S.
\end{acknowledgements}

\bibliographystyle{apsrev4-2} 
\bibliography{bib}

\ifdefined\DeclarePrefChars\DeclarePrefChars{'’-}\else\fi  \newcommand{\noop}[1]{}
\begin{thebibliography}{71}%
\makeatletter
\providecommand \@ifxundefined [1]{%
 \@ifx{#1\undefined}
}%
\providecommand \@ifnum [1]{%
 \ifnum #1\expandafter \@firstoftwo
 \else \expandafter \@secondoftwo
 \fi
}%
\providecommand \@ifx [1]{%
 \ifx #1\expandafter \@firstoftwo
 \else \expandafter \@secondoftwo
 \fi
}%
\providecommand \natexlab [1]{#1}%
\providecommand \enquote  [1]{``#1''}%
\providecommand \bibnamefont  [1]{#1}%
\providecommand \bibfnamefont [1]{#1}%
\providecommand \citenamefont [1]{#1}%
\providecommand \href@noop [0]{\@secondoftwo}%
\providecommand \href [0]{\begingroup \@sanitize@url \@href}%
\providecommand \@href[1]{\@@startlink{#1}\@@href}%
\providecommand \@@href[1]{\endgroup#1\@@endlink}%
\providecommand \@sanitize@url [0]{\catcode `\\12\catcode `\$12\catcode `\&12\catcode `\#12\catcode `\^12\catcode `\_12\catcode `\%12\relax}%
\providecommand \@@startlink[1]{}%
\providecommand \@@endlink[0]{}%
\providecommand \url  [0]{\begingroup\@sanitize@url \@url }%
\providecommand \@url [1]{\endgroup\@href {#1}{\urlprefix }}%
\providecommand \urlprefix  [0]{URL }%
\providecommand \Eprint [0]{\href }%
\providecommand \doibase [0]{https://doi.org/}%
\providecommand \selectlanguage [0]{\@gobble}%
\providecommand \bibinfo  [0]{\@secondoftwo}%
\providecommand \bibfield  [0]{\@secondoftwo}%
\providecommand \translation [1]{[#1]}%
\providecommand \BibitemOpen [0]{}%
\providecommand \bibitemStop [0]{}%
\providecommand \bibitemNoStop [0]{.\EOS\space}%
\providecommand \EOS [0]{\spacefactor3000\relax}%
\providecommand \BibitemShut  [1]{\csname bibitem#1\endcsname}%
\let\auto@bib@innerbib\@empty
\bibitem [{\citenamefont {Cai}\ \emph {et~al.}(2023)\citenamefont {Cai}, \citenamefont {Anderson}, \citenamefont {Wang}, \citenamefont {Zhang}, \citenamefont {Liu}, \citenamefont {Holtzmann}, \citenamefont {Zhang}, \citenamefont {Fan}, \citenamefont {Taniguchi}, \citenamefont {Watanabe}, \citenamefont {Ran}, \citenamefont {Cao}, \citenamefont {Fu}, \citenamefont {Xiao}, \citenamefont {Yao},\ and\ \citenamefont {Xu}}]{Cai_2023}%
  \BibitemOpen
  \bibfield  {author} {\bibinfo {author} {\bibfnamefont {J.}~\bibnamefont {Cai}}, \bibinfo {author} {\bibfnamefont {E.}~\bibnamefont {Anderson}}, \bibinfo {author} {\bibfnamefont {C.}~\bibnamefont {Wang}}, \bibinfo {author} {\bibfnamefont {X.}~\bibnamefont {Zhang}}, \bibinfo {author} {\bibfnamefont {X.}~\bibnamefont {Liu}}, \bibinfo {author} {\bibfnamefont {W.}~\bibnamefont {Holtzmann}}, \bibinfo {author} {\bibfnamefont {Y.}~\bibnamefont {Zhang}}, \bibinfo {author} {\bibfnamefont {F.}~\bibnamefont {Fan}}, \bibinfo {author} {\bibfnamefont {T.}~\bibnamefont {Taniguchi}}, \bibinfo {author} {\bibfnamefont {K.}~\bibnamefont {Watanabe}}, \bibinfo {author} {\bibfnamefont {Y.}~\bibnamefont {Ran}}, \bibinfo {author} {\bibfnamefont {T.}~\bibnamefont {Cao}}, \bibinfo {author} {\bibfnamefont {L.}~\bibnamefont {Fu}}, \bibinfo {author} {\bibfnamefont {D.}~\bibnamefont {Xiao}}, \bibinfo {author} {\bibfnamefont {W.}~\bibnamefont {Yao}},\ and\ \bibinfo {author} {\bibfnamefont {X.}~\bibnamefont {Xu}},\ }\href
  {https://doi.org/10.1038/s41586-023-06289-w} {\bibfield  {journal} {\bibinfo  {journal} {Nature}\ }\textbf {\bibinfo {volume} {622}},\ \bibinfo {pages} {63–68} (\bibinfo {year} {2023})}\BibitemShut {NoStop}%
\bibitem [{\citenamefont {Park}\ \emph {et~al.}(2023)\citenamefont {Park}, \citenamefont {Cai}, \citenamefont {Anderson}, \citenamefont {Zhang}, \citenamefont {Zhu}, \citenamefont {Liu}, \citenamefont {Wang}, \citenamefont {Holtzmann}, \citenamefont {Hu}, \citenamefont {Liu}, \citenamefont {Taniguchi}, \citenamefont {Watanabe}, \citenamefont {Chu}, \citenamefont {Cao}, \citenamefont {Fu}, \citenamefont {Yao}, \citenamefont {Chang}, \citenamefont {Cobden}, \citenamefont {Xiao},\ and\ \citenamefont {Xu}}]{Park2023}%
  \BibitemOpen
  \bibfield  {author} {\bibinfo {author} {\bibfnamefont {H.}~\bibnamefont {Park}}, \bibinfo {author} {\bibfnamefont {J.}~\bibnamefont {Cai}}, \bibinfo {author} {\bibfnamefont {E.}~\bibnamefont {Anderson}}, \bibinfo {author} {\bibfnamefont {Y.}~\bibnamefont {Zhang}}, \bibinfo {author} {\bibfnamefont {J.}~\bibnamefont {Zhu}}, \bibinfo {author} {\bibfnamefont {X.}~\bibnamefont {Liu}}, \bibinfo {author} {\bibfnamefont {C.}~\bibnamefont {Wang}}, \bibinfo {author} {\bibfnamefont {W.}~\bibnamefont {Holtzmann}}, \bibinfo {author} {\bibfnamefont {C.}~\bibnamefont {Hu}}, \bibinfo {author} {\bibfnamefont {Z.}~\bibnamefont {Liu}}, \bibinfo {author} {\bibfnamefont {T.}~\bibnamefont {Taniguchi}}, \bibinfo {author} {\bibfnamefont {K.}~\bibnamefont {Watanabe}}, \bibinfo {author} {\bibfnamefont {J.-H.}\ \bibnamefont {Chu}}, \bibinfo {author} {\bibfnamefont {T.}~\bibnamefont {Cao}}, \bibinfo {author} {\bibfnamefont {L.}~\bibnamefont {Fu}}, \bibinfo {author} {\bibfnamefont {W.}~\bibnamefont {Yao}}, \bibinfo {author}
  {\bibfnamefont {C.-Z.}\ \bibnamefont {Chang}}, \bibinfo {author} {\bibfnamefont {D.}~\bibnamefont {Cobden}}, \bibinfo {author} {\bibfnamefont {D.}~\bibnamefont {Xiao}},\ and\ \bibinfo {author} {\bibfnamefont {X.}~\bibnamefont {Xu}},\ }\href {https://doi.org/10.1038/s41586-023-06536-0} {\bibfield  {journal} {\bibinfo  {journal} {Nature}\ }\textbf {\bibinfo {volume} {622}},\ \bibinfo {pages} {74} (\bibinfo {year} {2023})}\BibitemShut {NoStop}%
\bibitem [{\citenamefont {Xu}\ \emph {et~al.}(2023)\citenamefont {Xu}, \citenamefont {Sun}, \citenamefont {Jia}, \citenamefont {Liu}, \citenamefont {Xu}, \citenamefont {Li}, \citenamefont {Gu}, \citenamefont {Watanabe}, \citenamefont {Taniguchi}, \citenamefont {Tong}, \citenamefont {Jia}, \citenamefont {Shi}, \citenamefont {Jiang}, \citenamefont {Zhang}, \citenamefont {Liu},\ and\ \citenamefont {Li}}]{Xu_2023}%
  \BibitemOpen
  \bibfield  {author} {\bibinfo {author} {\bibfnamefont {F.}~\bibnamefont {Xu}}, \bibinfo {author} {\bibfnamefont {Z.}~\bibnamefont {Sun}}, \bibinfo {author} {\bibfnamefont {T.}~\bibnamefont {Jia}}, \bibinfo {author} {\bibfnamefont {C.}~\bibnamefont {Liu}}, \bibinfo {author} {\bibfnamefont {C.}~\bibnamefont {Xu}}, \bibinfo {author} {\bibfnamefont {C.}~\bibnamefont {Li}}, \bibinfo {author} {\bibfnamefont {Y.}~\bibnamefont {Gu}}, \bibinfo {author} {\bibfnamefont {K.}~\bibnamefont {Watanabe}}, \bibinfo {author} {\bibfnamefont {T.}~\bibnamefont {Taniguchi}}, \bibinfo {author} {\bibfnamefont {B.}~\bibnamefont {Tong}}, \bibinfo {author} {\bibfnamefont {J.}~\bibnamefont {Jia}}, \bibinfo {author} {\bibfnamefont {Z.}~\bibnamefont {Shi}}, \bibinfo {author} {\bibfnamefont {S.}~\bibnamefont {Jiang}}, \bibinfo {author} {\bibfnamefont {Y.}~\bibnamefont {Zhang}}, \bibinfo {author} {\bibfnamefont {X.}~\bibnamefont {Liu}},\ and\ \bibinfo {author} {\bibfnamefont {T.}~\bibnamefont {Li}},\ }\bibfield  {journal} {\bibinfo
  {journal} {Physical Review X}\ }\textbf {\bibinfo {volume} {13}},\ \href {https://doi.org/10.1103/physrevx.13.031037} {10.1103/physrevx.13.031037} (\bibinfo {year} {2023})\BibitemShut {NoStop}%
\bibitem [{\citenamefont {Zeng}\ \emph {et~al.}(2023)\citenamefont {Zeng}, \citenamefont {Xia}, \citenamefont {Kang}, \citenamefont {Zhu}, \citenamefont {Knüppel}, \citenamefont {Vaswani}, \citenamefont {Watanabe}, \citenamefont {Taniguchi}, \citenamefont {Mak},\ and\ \citenamefont {Shan}}]{Zeng_2023}%
  \BibitemOpen
  \bibfield  {author} {\bibinfo {author} {\bibfnamefont {Y.}~\bibnamefont {Zeng}}, \bibinfo {author} {\bibfnamefont {Z.}~\bibnamefont {Xia}}, \bibinfo {author} {\bibfnamefont {K.}~\bibnamefont {Kang}}, \bibinfo {author} {\bibfnamefont {J.}~\bibnamefont {Zhu}}, \bibinfo {author} {\bibfnamefont {P.}~\bibnamefont {Knüppel}}, \bibinfo {author} {\bibfnamefont {C.}~\bibnamefont {Vaswani}}, \bibinfo {author} {\bibfnamefont {K.}~\bibnamefont {Watanabe}}, \bibinfo {author} {\bibfnamefont {T.}~\bibnamefont {Taniguchi}}, \bibinfo {author} {\bibfnamefont {K.~F.}\ \bibnamefont {Mak}},\ and\ \bibinfo {author} {\bibfnamefont {J.}~\bibnamefont {Shan}},\ }\href {https://doi.org/10.1038/s41586-023-06452-3} {\bibfield  {journal} {\bibinfo  {journal} {Nature}\ }\textbf {\bibinfo {volume} {622}},\ \bibinfo {pages} {69–73} (\bibinfo {year} {2023})}\BibitemShut {NoStop}%
\bibitem [{\citenamefont {Lu}\ \emph {et~al.}(2024)\citenamefont {Lu}, \citenamefont {Han}, \citenamefont {Yao}, \citenamefont {Reddy}, \citenamefont {Yang}, \citenamefont {Seo}, \citenamefont {Watanabe}, \citenamefont {Taniguchi}, \citenamefont {Fu},\ and\ \citenamefont {Ju}}]{Lu2024}%
  \BibitemOpen
  \bibfield  {author} {\bibinfo {author} {\bibfnamefont {Z.}~\bibnamefont {Lu}}, \bibinfo {author} {\bibfnamefont {T.}~\bibnamefont {Han}}, \bibinfo {author} {\bibfnamefont {Y.}~\bibnamefont {Yao}}, \bibinfo {author} {\bibfnamefont {A.~P.}\ \bibnamefont {Reddy}}, \bibinfo {author} {\bibfnamefont {J.}~\bibnamefont {Yang}}, \bibinfo {author} {\bibfnamefont {J.}~\bibnamefont {Seo}}, \bibinfo {author} {\bibfnamefont {K.}~\bibnamefont {Watanabe}}, \bibinfo {author} {\bibfnamefont {T.}~\bibnamefont {Taniguchi}}, \bibinfo {author} {\bibfnamefont {L.}~\bibnamefont {Fu}},\ and\ \bibinfo {author} {\bibfnamefont {L.}~\bibnamefont {Ju}},\ }\href {https://doi.org/10.1038/s41586-023-07010-7} {\bibfield  {journal} {\bibinfo  {journal} {Nature}\ }\textbf {\bibinfo {volume} {626}},\ \bibinfo {pages} {759} (\bibinfo {year} {2024})}\BibitemShut {NoStop}%
\bibitem [{\citenamefont {Kitaev}(2003)}]{kitaev2003fault}%
  \BibitemOpen
  \bibfield  {author} {\bibinfo {author} {\bibfnamefont {A.~Y.}\ \bibnamefont {Kitaev}},\ }\href@noop {} {\bibfield  {journal} {\bibinfo  {journal} {Annals of physics}\ }\textbf {\bibinfo {volume} {303}},\ \bibinfo {pages} {2} (\bibinfo {year} {2003})}\BibitemShut {NoStop}%
\bibitem [{\citenamefont {Liu}\ and\ \citenamefont {Bergholtz}(2022)}]{liuRecentDevelopmentsFractional2022}%
  \BibitemOpen
  \bibfield  {author} {\bibinfo {author} {\bibfnamefont {Z.}~\bibnamefont {Liu}}\ and\ \bibinfo {author} {\bibfnamefont {E.~J.}\ \bibnamefont {Bergholtz}},\ }\href {https://doi.org/10.48550/arXiv.2208.08449} {\bibinfo {title} {Recent {{Developments}} in {{Fractional Chern Insulators}}}} (\bibinfo {year} {2022}),\ \Eprint {https://arxiv.org/abs/2208.08449} {arXiv:2208.08449 [cond-mat, physics:math-ph, physics:quant-ph]} \BibitemShut {NoStop}%
\bibitem [{\citenamefont {Parameswaran}\ \emph {et~al.}(2013)\citenamefont {Parameswaran}, \citenamefont {Roy},\ and\ \citenamefont {Sondhi}}]{parameswaranFractionalQuantumHall2013}%
  \BibitemOpen
  \bibfield  {author} {\bibinfo {author} {\bibfnamefont {S.~A.}\ \bibnamefont {Parameswaran}}, \bibinfo {author} {\bibfnamefont {R.}~\bibnamefont {Roy}},\ and\ \bibinfo {author} {\bibfnamefont {S.~L.}\ \bibnamefont {Sondhi}},\ }\href {https://doi.org/10.1016/j.crhy.2013.04.003} {\bibfield  {journal} {\bibinfo  {journal} {C. R. Phys.}\ }\textbf {\bibinfo {volume} {14}},\ \bibinfo {pages} {816} (\bibinfo {year} {2013})}\BibitemShut {NoStop}%
\bibitem [{\citenamefont {Bergholtz}\ and\ \citenamefont {Liu}(2013)}]{BergholtzReview2013}%
  \BibitemOpen
  \bibfield  {author} {\bibinfo {author} {\bibfnamefont {E.~J.}\ \bibnamefont {Bergholtz}}\ and\ \bibinfo {author} {\bibfnamefont {Z.}~\bibnamefont {Liu}},\ }\href {https://doi.org/10.1142/S021797921330017X} {\bibfield  {journal} {\bibinfo  {journal} {Int. J. Mod. Phys. B}\ }\textbf {\bibinfo {volume} {27}},\ \bibinfo {pages} {1330017} (\bibinfo {year} {2013})},\ \Eprint {https://arxiv.org/abs/https://doi.org/10.1142/S021797921330017X} {https://doi.org/10.1142/S021797921330017X} \BibitemShut {NoStop}%
\bibitem [{\citenamefont {Neupert}\ \emph {et~al.}(2011)\citenamefont {Neupert}, \citenamefont {Santos}, \citenamefont {Chamon},\ and\ \citenamefont {Mudry}}]{neupertFractionalQuantumHall2011}%
  \BibitemOpen
  \bibfield  {author} {\bibinfo {author} {\bibfnamefont {T.}~\bibnamefont {Neupert}}, \bibinfo {author} {\bibfnamefont {L.}~\bibnamefont {Santos}}, \bibinfo {author} {\bibfnamefont {C.}~\bibnamefont {Chamon}},\ and\ \bibinfo {author} {\bibfnamefont {C.}~\bibnamefont {Mudry}},\ }\href {https://doi.org/10.1103/PhysRevLett.106.236804} {\bibfield  {journal} {\bibinfo  {journal} {Phys. Rev. Lett.}\ }\textbf {\bibinfo {volume} {106}},\ \bibinfo {pages} {236804} (\bibinfo {year} {2011})}\BibitemShut {NoStop}%
\bibitem [{\citenamefont {Sheng}\ \emph {et~al.}(2011)\citenamefont {Sheng}, \citenamefont {Gu}, \citenamefont {Sun},\ and\ \citenamefont {Sheng}}]{shengFractionalQuantumHall2011}%
  \BibitemOpen
  \bibfield  {author} {\bibinfo {author} {\bibfnamefont {D.}~\bibnamefont {Sheng}}, \bibinfo {author} {\bibfnamefont {Z.-C.}\ \bibnamefont {Gu}}, \bibinfo {author} {\bibfnamefont {K.}~\bibnamefont {Sun}},\ and\ \bibinfo {author} {\bibfnamefont {L.}~\bibnamefont {Sheng}},\ }\href {https://doi.org/10.1038/ncomms1380} {\bibfield  {journal} {\bibinfo  {journal} {Nat. Commun.}\ }\textbf {\bibinfo {volume} {2}},\ \bibinfo {pages} {389} (\bibinfo {year} {2011})}\BibitemShut {NoStop}%
\bibitem [{\citenamefont {Regnault}\ and\ \citenamefont {Bernevig}(2011)}]{regnaultFractionalChernInsulator2011}%
  \BibitemOpen
  \bibfield  {author} {\bibinfo {author} {\bibfnamefont {N.}~\bibnamefont {Regnault}}\ and\ \bibinfo {author} {\bibfnamefont {B.~A.}\ \bibnamefont {Bernevig}},\ }\href {https://doi.org/10.1103/PhysRevX.1.021014} {\bibfield  {journal} {\bibinfo  {journal} {Phys. Rev. X}\ }\textbf {\bibinfo {volume} {1}},\ \bibinfo {pages} {021014} (\bibinfo {year} {2011})}\BibitemShut {NoStop}%
\bibitem [{\citenamefont {Qi}(2011)}]{qi_generic_2011}%
  \BibitemOpen
  \bibfield  {author} {\bibinfo {author} {\bibfnamefont {X.-L.}\ \bibnamefont {Qi}},\ }\href {https://doi.org/10.1103/PhysRevLett.107.126803} {\bibfield  {journal} {\bibinfo  {journal} {Phys. Rev. Lett.}\ }\textbf {\bibinfo {volume} {107}},\ \bibinfo {pages} {126803} (\bibinfo {year} {2011})}\BibitemShut {NoStop}%
\bibitem [{\citenamefont {Parameswaran}\ \emph {et~al.}(2012)\citenamefont {Parameswaran}, \citenamefont {Roy},\ and\ \citenamefont {Sondhi}}]{parameswaranFractionalChernInsulators2012}%
  \BibitemOpen
  \bibfield  {author} {\bibinfo {author} {\bibfnamefont {S.~A.}\ \bibnamefont {Parameswaran}}, \bibinfo {author} {\bibfnamefont {R.}~\bibnamefont {Roy}},\ and\ \bibinfo {author} {\bibfnamefont {S.~L.}\ \bibnamefont {Sondhi}},\ }\href {https://doi.org/10.1103/PhysRevB.85.241308} {\bibfield  {journal} {\bibinfo  {journal} {Phys. Rev. B}\ }\textbf {\bibinfo {volume} {85}},\ \bibinfo {pages} {241308} (\bibinfo {year} {2012})}\BibitemShut {NoStop}%
\bibitem [{\citenamefont {Wu}\ \emph {et~al.}(2013)\citenamefont {Wu}, \citenamefont {Regnault},\ and\ \citenamefont {Bernevig}}]{wuBlochModelWave2013}%
  \BibitemOpen
  \bibfield  {author} {\bibinfo {author} {\bibfnamefont {Y.-L.}\ \bibnamefont {Wu}}, \bibinfo {author} {\bibfnamefont {N.}~\bibnamefont {Regnault}},\ and\ \bibinfo {author} {\bibfnamefont {B.~A.}\ \bibnamefont {Bernevig}},\ }\href {https://doi.org/10.1103/PhysRevLett.110.106802} {\bibfield  {journal} {\bibinfo  {journal} {Phys. Rev. Lett.}\ }\textbf {\bibinfo {volume} {110}},\ \bibinfo {pages} {106802} (\bibinfo {year} {2013})}\BibitemShut {NoStop}%
\bibitem [{\citenamefont {Kourtis}\ \emph {et~al.}(2014)\citenamefont {Kourtis}, \citenamefont {Neupert}, \citenamefont {Chamon},\ and\ \citenamefont {Mudry}}]{kourtisFractionalChernInsulators2014}%
  \BibitemOpen
  \bibfield  {author} {\bibinfo {author} {\bibfnamefont {S.}~\bibnamefont {Kourtis}}, \bibinfo {author} {\bibfnamefont {T.}~\bibnamefont {Neupert}}, \bibinfo {author} {\bibfnamefont {C.}~\bibnamefont {Chamon}},\ and\ \bibinfo {author} {\bibfnamefont {C.}~\bibnamefont {Mudry}},\ }\href {https://doi.org/10.1103/PhysRevLett.112.126806} {\bibfield  {journal} {\bibinfo  {journal} {Physical Review Letters}\ }\textbf {\bibinfo {volume} {112}},\ \bibinfo {pages} {126806} (\bibinfo {year} {2014})},\ \Eprint {https://arxiv.org/abs/1310.6371} {arXiv:1310.6371 [cond-mat]} \BibitemShut {NoStop}%
\bibitem [{\citenamefont {Zhang}\ \emph {et~al.}(2019)\citenamefont {Zhang}, \citenamefont {Mao}, \citenamefont {Cao}, \citenamefont {Jarillo-Herrero},\ and\ \citenamefont {Senthil}}]{YahuiChern}%
  \BibitemOpen
  \bibfield  {author} {\bibinfo {author} {\bibfnamefont {Y.-H.}\ \bibnamefont {Zhang}}, \bibinfo {author} {\bibfnamefont {D.}~\bibnamefont {Mao}}, \bibinfo {author} {\bibfnamefont {Y.}~\bibnamefont {Cao}}, \bibinfo {author} {\bibfnamefont {P.}~\bibnamefont {Jarillo-Herrero}},\ and\ \bibinfo {author} {\bibfnamefont {T.}~\bibnamefont {Senthil}},\ }\href {https://doi.org/10.1103/PhysRevB.99.075127} {\bibfield  {journal} {\bibinfo  {journal} {Phys. Rev. B}\ }\textbf {\bibinfo {volume} {99}},\ \bibinfo {pages} {075127} (\bibinfo {year} {2019})}\BibitemShut {NoStop}%
\bibitem [{\citenamefont {Tarnopolsky}\ \emph {et~al.}(2019)\citenamefont {Tarnopolsky}, \citenamefont {Kruchkov},\ and\ \citenamefont {Vishwanath}}]{tarnopolskyOriginMagicAngles2019}%
  \BibitemOpen
  \bibfield  {author} {\bibinfo {author} {\bibfnamefont {G.}~\bibnamefont {Tarnopolsky}}, \bibinfo {author} {\bibfnamefont {A.~J.}\ \bibnamefont {Kruchkov}},\ and\ \bibinfo {author} {\bibfnamefont {A.}~\bibnamefont {Vishwanath}},\ }\href {https://doi.org/10.1103/PhysRevLett.122.106405} {\bibfield  {journal} {\bibinfo  {journal} {Phys. Rev. Lett.}\ }\textbf {\bibinfo {volume} {122}},\ \bibinfo {pages} {106405} (\bibinfo {year} {2019})}\BibitemShut {NoStop}%
\bibitem [{\citenamefont {Ledwith}\ \emph {et~al.}(2020{\natexlab{a}})\citenamefont {Ledwith}, \citenamefont {Tarnopolsky}, \citenamefont {Khalaf},\ and\ \citenamefont {Vishwanath}}]{ledwithFractionalChernInsulator2020a}%
  \BibitemOpen
  \bibfield  {author} {\bibinfo {author} {\bibfnamefont {P.~J.}\ \bibnamefont {Ledwith}}, \bibinfo {author} {\bibfnamefont {G.}~\bibnamefont {Tarnopolsky}}, \bibinfo {author} {\bibfnamefont {E.}~\bibnamefont {Khalaf}},\ and\ \bibinfo {author} {\bibfnamefont {A.}~\bibnamefont {Vishwanath}},\ }\href {https://doi.org/10.1103/PhysRevResearch.2.023237} {\bibfield  {journal} {\bibinfo  {journal} {Phys. Rev. Research}\ }\textbf {\bibinfo {volume} {2}},\ \bibinfo {pages} {023237} (\bibinfo {year} {2020}{\natexlab{a}})}\BibitemShut {NoStop}%
\bibitem [{\citenamefont {Repellin}\ and\ \citenamefont {Senthil}(2020)}]{CecilleFCI}%
  \BibitemOpen
  \bibfield  {author} {\bibinfo {author} {\bibfnamefont {C.}~\bibnamefont {Repellin}}\ and\ \bibinfo {author} {\bibfnamefont {T.}~\bibnamefont {Senthil}},\ }\href {https://doi.org/10.1103/PhysRevResearch.2.023238} {\bibfield  {journal} {\bibinfo  {journal} {Phys. Rev. Res.}\ }\textbf {\bibinfo {volume} {2}},\ \bibinfo {pages} {023238} (\bibinfo {year} {2020})}\BibitemShut {NoStop}%
\bibitem [{\citenamefont {Abouelkomsan}\ \emph {et~al.}(2020)\citenamefont {Abouelkomsan}, \citenamefont {Liu},\ and\ \citenamefont {Bergholtz}}]{AbouelkomsanTBG}%
  \BibitemOpen
  \bibfield  {author} {\bibinfo {author} {\bibfnamefont {A.}~\bibnamefont {Abouelkomsan}}, \bibinfo {author} {\bibfnamefont {Z.}~\bibnamefont {Liu}},\ and\ \bibinfo {author} {\bibfnamefont {E.~J.}\ \bibnamefont {Bergholtz}},\ }\href {https://doi.org/10.1103/PhysRevLett.124.106803} {\bibfield  {journal} {\bibinfo  {journal} {Phys. Rev. Lett.}\ }\textbf {\bibinfo {volume} {124}},\ \bibinfo {pages} {106803} (\bibinfo {year} {2020})}\BibitemShut {NoStop}%
\bibitem [{\citenamefont {Liu}\ \emph {et~al.}(2021)\citenamefont {Liu}, \citenamefont {Abouelkomsan},\ and\ \citenamefont {Bergholtz}}]{AbouelkomsanTDBG}%
  \BibitemOpen
  \bibfield  {author} {\bibinfo {author} {\bibfnamefont {Z.}~\bibnamefont {Liu}}, \bibinfo {author} {\bibfnamefont {A.}~\bibnamefont {Abouelkomsan}},\ and\ \bibinfo {author} {\bibfnamefont {E.~J.}\ \bibnamefont {Bergholtz}},\ }\href {https://doi.org/10.1103/PhysRevLett.126.026801} {\bibfield  {journal} {\bibinfo  {journal} {Phys. Rev. Lett.}\ }\textbf {\bibinfo {volume} {126}},\ \bibinfo {pages} {026801} (\bibinfo {year} {2021})}\BibitemShut {NoStop}%
\bibitem [{\citenamefont {Mera}\ and\ \citenamefont {Ozawa}(2021{\natexlab{a}})}]{meraEngineeringGeometricallyFlat2021}%
  \BibitemOpen
  \bibfield  {author} {\bibinfo {author} {\bibfnamefont {B.}~\bibnamefont {Mera}}\ and\ \bibinfo {author} {\bibfnamefont {T.}~\bibnamefont {Ozawa}},\ }\href {https://doi.org/10.1103/PhysRevB.104.115160} {\bibfield  {journal} {\bibinfo  {journal} {Phys. Rev. B}\ }\textbf {\bibinfo {volume} {104}},\ \bibinfo {pages} {115160} (\bibinfo {year} {2021}{\natexlab{a}})}\BibitemShut {NoStop}%
\bibitem [{\citenamefont {Wang}\ \emph {et~al.}(2021)\citenamefont {Wang}, \citenamefont {Cano}, \citenamefont {Millis}, \citenamefont {Liu},\ and\ \citenamefont {Yang}}]{Wang_2021}%
  \BibitemOpen
  \bibfield  {author} {\bibinfo {author} {\bibfnamefont {J.}~\bibnamefont {Wang}}, \bibinfo {author} {\bibfnamefont {J.}~\bibnamefont {Cano}}, \bibinfo {author} {\bibfnamefont {A.~J.}\ \bibnamefont {Millis}}, \bibinfo {author} {\bibfnamefont {Z.}~\bibnamefont {Liu}},\ and\ \bibinfo {author} {\bibfnamefont {B.}~\bibnamefont {Yang}},\ }\href {https://doi.org/10.1103/PhysRevLett.127.246403} {\bibfield  {journal} {\bibinfo  {journal} {Phys. Rev. Lett.}\ }\textbf {\bibinfo {volume} {127}},\ \bibinfo {pages} {246403} (\bibinfo {year} {2021})}\BibitemShut {NoStop}%
\bibitem [{\citenamefont {Ledwith}\ \emph {et~al.}(2022{\natexlab{a}})\citenamefont {Ledwith}, \citenamefont {Vishwanath},\ and\ \citenamefont {Khalaf}}]{ledwithFamilyIdealChern2022}%
  \BibitemOpen
  \bibfield  {author} {\bibinfo {author} {\bibfnamefont {P.~J.}\ \bibnamefont {Ledwith}}, \bibinfo {author} {\bibfnamefont {A.}~\bibnamefont {Vishwanath}},\ and\ \bibinfo {author} {\bibfnamefont {E.}~\bibnamefont {Khalaf}},\ }\href {https://doi.org/10.1103/PhysRevLett.128.176404} {\bibfield  {journal} {\bibinfo  {journal} {Phys. Rev. Lett.}\ }\textbf {\bibinfo {volume} {128}},\ \bibinfo {pages} {176404} (\bibinfo {year} {2022}{\natexlab{a}})}\BibitemShut {NoStop}%
\bibitem [{\citenamefont {Ledwith}\ \emph {et~al.}(2022{\natexlab{b}})\citenamefont {Ledwith}, \citenamefont {Vishwanath},\ and\ \citenamefont {Parker}}]{vortexability}%
  \BibitemOpen
  \bibfield  {author} {\bibinfo {author} {\bibfnamefont {P.~J.}\ \bibnamefont {Ledwith}}, \bibinfo {author} {\bibfnamefont {A.}~\bibnamefont {Vishwanath}},\ and\ \bibinfo {author} {\bibfnamefont {D.~E.}\ \bibnamefont {Parker}},\ }\href {https://arxiv.org/abs/2209.15023} {\  (\bibinfo {year} {2022}{\natexlab{b}})},\ \Eprint {https://arxiv.org/abs/arXiv: 2209.15023} {arXiv: 2209.15023} \BibitemShut {NoStop}%
\bibitem [{\citenamefont {Shi}\ and\ \citenamefont {Senthil}(2024)}]{shi2024}%
  \BibitemOpen
  \bibfield  {author} {\bibinfo {author} {\bibfnamefont {Z.~D.}\ \bibnamefont {Shi}}\ and\ \bibinfo {author} {\bibfnamefont {T.}~\bibnamefont {Senthil}},\ }\href {https://arxiv.org/abs/2409.20567} {\bibinfo {title} {Doping a fractional quantum anomalous hall insulator}} (\bibinfo {year} {2024}),\ \Eprint {https://arxiv.org/abs/2409.20567} {arXiv:2409.20567 [cond-mat.str-el]} \BibitemShut {NoStop}%
\bibitem [{\citenamefont {Gon{\c{c}}alves}\ \emph {et~al.}(2025)\citenamefont {Gon{\c{c}}alves}, \citenamefont {Mendez-Valderrama}, \citenamefont {Herzog-Arbeitman}, \citenamefont {Yu}, \citenamefont {Xu}, \citenamefont {Xiao}, \citenamefont {Bernevig},\ and\ \citenamefont {Regnault}}]{RegnaultDispersion}%
  \BibitemOpen
  \bibfield  {author} {\bibinfo {author} {\bibfnamefont {M.}~\bibnamefont {Gon{\c{c}}alves}}, \bibinfo {author} {\bibfnamefont {J.~F.}\ \bibnamefont {Mendez-Valderrama}}, \bibinfo {author} {\bibfnamefont {J.}~\bibnamefont {Herzog-Arbeitman}}, \bibinfo {author} {\bibfnamefont {J.}~\bibnamefont {Yu}}, \bibinfo {author} {\bibfnamefont {X.}~\bibnamefont {Xu}}, \bibinfo {author} {\bibfnamefont {D.}~\bibnamefont {Xiao}}, \bibinfo {author} {\bibfnamefont {B.~A.}\ \bibnamefont {Bernevig}},\ and\ \bibinfo {author} {\bibfnamefont {N.}~\bibnamefont {Regnault}},\ }\href@noop {} {\bibfield  {journal} {\bibinfo  {journal} {arXiv preprint arXiv:2506.05330}\ } (\bibinfo {year} {2025})}\BibitemShut {NoStop}%
\bibitem [{\citenamefont {Laughlin}(1988)}]{Laughlin_1988}%
  \BibitemOpen
  \bibfield  {author} {\bibinfo {author} {\bibfnamefont {R.~B.}\ \bibnamefont {Laughlin}},\ }\href {https://doi.org/10.1103/PhysRevLett.60.2677} {\bibfield  {journal} {\bibinfo  {journal} {Phys. Rev. Lett.}\ }\textbf {\bibinfo {volume} {60}},\ \bibinfo {pages} {2677} (\bibinfo {year} {1988})}\BibitemShut {NoStop}%
\bibitem [{\citenamefont {Fetter}\ \emph {et~al.}(1989)\citenamefont {Fetter}, \citenamefont {Hanna},\ and\ \citenamefont {Laughlin}}]{Fetter_1989}%
  \BibitemOpen
  \bibfield  {author} {\bibinfo {author} {\bibfnamefont {A.~L.}\ \bibnamefont {Fetter}}, \bibinfo {author} {\bibfnamefont {C.~B.}\ \bibnamefont {Hanna}},\ and\ \bibinfo {author} {\bibfnamefont {R.~B.}\ \bibnamefont {Laughlin}},\ }\href {https://doi.org/10.1103/PhysRevB.39.9679} {\bibfield  {journal} {\bibinfo  {journal} {Phys. Rev. B}\ }\textbf {\bibinfo {volume} {39}},\ \bibinfo {pages} {9679} (\bibinfo {year} {1989})}\BibitemShut {NoStop}%
\bibitem [{\citenamefont {Lee}\ and\ \citenamefont {Fisher}(1989)}]{Lee_1989}%
  \BibitemOpen
  \bibfield  {author} {\bibinfo {author} {\bibfnamefont {D.-H.}\ \bibnamefont {Lee}}\ and\ \bibinfo {author} {\bibfnamefont {M.~P.~A.}\ \bibnamefont {Fisher}},\ }\href {https://doi.org/10.1103/PhysRevLett.63.903} {\bibfield  {journal} {\bibinfo  {journal} {Phys. Rev. Lett.}\ }\textbf {\bibinfo {volume} {63}},\ \bibinfo {pages} {903} (\bibinfo {year} {1989})}\BibitemShut {NoStop}%
\bibitem [{\citenamefont {CHEN}\ \emph {et~al.}(1989)\citenamefont {CHEN}, \citenamefont {WILCZEK}, \citenamefont {WITTEN},\ and\ \citenamefont {HALPERIN}}]{Chen_1989}%
  \BibitemOpen
  \bibfield  {author} {\bibinfo {author} {\bibfnamefont {Y.-H.}\ \bibnamefont {CHEN}}, \bibinfo {author} {\bibfnamefont {F.}~\bibnamefont {WILCZEK}}, \bibinfo {author} {\bibfnamefont {E.}~\bibnamefont {WITTEN}},\ and\ \bibinfo {author} {\bibfnamefont {B.~I.}\ \bibnamefont {HALPERIN}},\ }\href {https://doi.org/10.1142/S0217979289000725} {\bibfield  {journal} {\bibinfo  {journal} {International Journal of Modern Physics B}\ }\textbf {\bibinfo {volume} {03}},\ \bibinfo {pages} {1001} (\bibinfo {year} {1989})},\ \Eprint {https://arxiv.org/abs/https://doi.org/10.1142/S0217979289000725} {https://doi.org/10.1142/S0217979289000725} \BibitemShut {NoStop}%
\bibitem [{\citenamefont {Halperin}\ \emph {et~al.}(1989)\citenamefont {Halperin}, \citenamefont {March-Russell},\ and\ \citenamefont {Wilczek}}]{Halperin_1989}%
  \BibitemOpen
  \bibfield  {author} {\bibinfo {author} {\bibfnamefont {B.~I.}\ \bibnamefont {Halperin}}, \bibinfo {author} {\bibfnamefont {J.}~\bibnamefont {March-Russell}},\ and\ \bibinfo {author} {\bibfnamefont {F.}~\bibnamefont {Wilczek}},\ }\href {https://doi.org/10.1103/PhysRevB.40.8726} {\bibfield  {journal} {\bibinfo  {journal} {Phys. Rev. B}\ }\textbf {\bibinfo {volume} {40}},\ \bibinfo {pages} {8726} (\bibinfo {year} {1989})}\BibitemShut {NoStop}%
\bibitem [{\citenamefont {Wen}\ and\ \citenamefont {Zee}(1990)}]{Wen_1990}%
  \BibitemOpen
  \bibfield  {author} {\bibinfo {author} {\bibfnamefont {X.~G.}\ \bibnamefont {Wen}}\ and\ \bibinfo {author} {\bibfnamefont {A.}~\bibnamefont {Zee}},\ }\href {https://doi.org/10.1103/PhysRevB.41.240} {\bibfield  {journal} {\bibinfo  {journal} {Phys. Rev. B}\ }\textbf {\bibinfo {volume} {41}},\ \bibinfo {pages} {240} (\bibinfo {year} {1990})}\BibitemShut {NoStop}%
\bibitem [{\citenamefont {Wen}\ and\ \citenamefont {Zee}(1991)}]{Wen_1991}%
  \BibitemOpen
  \bibfield  {author} {\bibinfo {author} {\bibfnamefont {X.~G.}\ \bibnamefont {Wen}}\ and\ \bibinfo {author} {\bibfnamefont {A.}~\bibnamefont {Zee}},\ }\href {https://doi.org/10.1103/PhysRevB.44.274} {\bibfield  {journal} {\bibinfo  {journal} {Phys. Rev. B}\ }\textbf {\bibinfo {volume} {44}},\ \bibinfo {pages} {274} (\bibinfo {year} {1991})}\BibitemShut {NoStop}%
\bibitem [{\citenamefont {Tang}\ and\ \citenamefont {Wen}(2013)}]{Tang_2013}%
  \BibitemOpen
  \bibfield  {author} {\bibinfo {author} {\bibfnamefont {E.}~\bibnamefont {Tang}}\ and\ \bibinfo {author} {\bibfnamefont {X.-G.}\ \bibnamefont {Wen}},\ }\href {https://doi.org/10.1103/PhysRevB.88.195117} {\bibfield  {journal} {\bibinfo  {journal} {Phys. Rev. B}\ }\textbf {\bibinfo {volume} {88}},\ \bibinfo {pages} {195117} (\bibinfo {year} {2013})}\BibitemShut {NoStop}%
\bibitem [{\citenamefont {Kim}\ \emph {et~al.}(2025)\citenamefont {Kim}, \citenamefont {Timmel}, \citenamefont {Ju},\ and\ \citenamefont {Wen}}]{kim2025topological}%
  \BibitemOpen
  \bibfield  {author} {\bibinfo {author} {\bibfnamefont {M.}~\bibnamefont {Kim}}, \bibinfo {author} {\bibfnamefont {A.}~\bibnamefont {Timmel}}, \bibinfo {author} {\bibfnamefont {L.}~\bibnamefont {Ju}},\ and\ \bibinfo {author} {\bibfnamefont {X.-G.}\ \bibnamefont {Wen}},\ }\href@noop {} {\bibfield  {journal} {\bibinfo  {journal} {Physical Review B}\ }\textbf {\bibinfo {volume} {111}},\ \bibinfo {pages} {014508} (\bibinfo {year} {2025})}\BibitemShut {NoStop}%
\bibitem [{\citenamefont {Divic}\ \emph {et~al.}(2024)\citenamefont {Divic}, \citenamefont {Cr\'epel}, \citenamefont {Soejima}, \citenamefont {Song}, \citenamefont {Millis}, \citenamefont {Zaletel},\ and\ \citenamefont {Vishwanath}}]{divic_2024}%
  \BibitemOpen
  \bibfield  {author} {\bibinfo {author} {\bibfnamefont {S.}~\bibnamefont {Divic}}, \bibinfo {author} {\bibfnamefont {V.}~\bibnamefont {Cr\'epel}}, \bibinfo {author} {\bibfnamefont {T.}~\bibnamefont {Soejima}}, \bibinfo {author} {\bibfnamefont {X.-Y.}\ \bibnamefont {Song}}, \bibinfo {author} {\bibfnamefont {A.}~\bibnamefont {Millis}}, \bibinfo {author} {\bibfnamefont {M.~P.}\ \bibnamefont {Zaletel}},\ and\ \bibinfo {author} {\bibfnamefont {A.}~\bibnamefont {Vishwanath}},\ }\href {https://arxiv.org/abs/2410.18175} {\  (\bibinfo {year} {2024})},\ \Eprint {https://arxiv.org/abs/2410.18175} {arXiv:2410.18175 [cond-mat.str-el]} \BibitemShut {NoStop}%
\bibitem [{\citenamefont {Shi}\ \emph {et~al.}(2025)\citenamefont {Shi}, \citenamefont {Zhang},\ and\ \citenamefont {Senthil}}]{shi2025dopinglatticenonabelianquantum}%
  \BibitemOpen
  \bibfield  {author} {\bibinfo {author} {\bibfnamefont {Z.~D.}\ \bibnamefont {Shi}}, \bibinfo {author} {\bibfnamefont {C.}~\bibnamefont {Zhang}},\ and\ \bibinfo {author} {\bibfnamefont {T.}~\bibnamefont {Senthil}},\ }\href {https://arxiv.org/abs/2505.02893} {\bibinfo {title} {Doping lattice non-abelian quantum hall states}} (\bibinfo {year} {2025}),\ \Eprint {https://arxiv.org/abs/2505.02893} {arXiv:2505.02893 [cond-mat.str-el]} \BibitemShut {NoStop}%
\bibitem [{\citenamefont {Shi}\ and\ \citenamefont {Senthil}(2025)}]{shi2025anyon}%
  \BibitemOpen
  \bibfield  {author} {\bibinfo {author} {\bibfnamefont {Z.~D.}\ \bibnamefont {Shi}}\ and\ \bibinfo {author} {\bibfnamefont {T.}~\bibnamefont {Senthil}},\ }\href@noop {} {\bibfield  {journal} {\bibinfo  {journal} {arXiv preprint arXiv:2506.02128}\ } (\bibinfo {year} {2025})}\BibitemShut {NoStop}%
\bibitem [{\citenamefont {Nosov}\ \emph {et~al.}(2025)\citenamefont {Nosov}, \citenamefont {Han},\ and\ \citenamefont {Khalaf}}]{nosov2025anyonsuperconductivityplateautransitions}%
  \BibitemOpen
  \bibfield  {author} {\bibinfo {author} {\bibfnamefont {P.~A.}\ \bibnamefont {Nosov}}, \bibinfo {author} {\bibfnamefont {Z.}~\bibnamefont {Han}},\ and\ \bibinfo {author} {\bibfnamefont {E.}~\bibnamefont {Khalaf}},\ }\href {https://arxiv.org/abs/2506.02108} {\bibinfo {title} {Anyon superconductivity and plateau transitions in doped fractional quantum anomalous hall insulators}} (\bibinfo {year} {2025}),\ \Eprint {https://arxiv.org/abs/2506.02108} {arXiv:2506.02108 [cond-mat.str-el]} \BibitemShut {NoStop}%
\bibitem [{\citenamefont {Pichler}\ \emph {et~al.}(2025)\citenamefont {Pichler}, \citenamefont {Kuhlenkamp}, \citenamefont {Knap},\ and\ \citenamefont {Vishwanath}}]{ClemensAnyonSC}%
  \BibitemOpen
  \bibfield  {author} {\bibinfo {author} {\bibfnamefont {F.}~\bibnamefont {Pichler}}, \bibinfo {author} {\bibfnamefont {C.}~\bibnamefont {Kuhlenkamp}}, \bibinfo {author} {\bibfnamefont {M.}~\bibnamefont {Knap}},\ and\ \bibinfo {author} {\bibfnamefont {A.}~\bibnamefont {Vishwanath}},\ }\href@noop {} {\bibfield  {journal} {\bibinfo  {journal} {arXiv preprint arXiv:2506.08000}\ } (\bibinfo {year} {2025})}\BibitemShut {NoStop}%
\bibitem [{\citenamefont {Kousa}\ \emph {et~al.}(2025)\citenamefont {Kousa}, \citenamefont {Morales-Dur{\'a}n}, \citenamefont {Wolf}, \citenamefont {Khalaf},\ and\ \citenamefont {MacDonald}}]{kousa2025theory}%
  \BibitemOpen
  \bibfield  {author} {\bibinfo {author} {\bibfnamefont {B.~M.}\ \bibnamefont {Kousa}}, \bibinfo {author} {\bibfnamefont {N.}~\bibnamefont {Morales-Dur{\'a}n}}, \bibinfo {author} {\bibfnamefont {T.~M.}\ \bibnamefont {Wolf}}, \bibinfo {author} {\bibfnamefont {E.}~\bibnamefont {Khalaf}},\ and\ \bibinfo {author} {\bibfnamefont {A.~H.}\ \bibnamefont {MacDonald}},\ }\href@noop {} {\bibfield  {journal} {\bibinfo  {journal} {arXiv preprint arXiv:2502.17574}\ } (\bibinfo {year} {2025})}\BibitemShut {NoStop}%
\bibitem [{\citenamefont {Reddy}\ \emph {et~al.}(2023)\citenamefont {Reddy}, \citenamefont {Alsallom}, \citenamefont {Zhang}, \citenamefont {Devakul},\ and\ \citenamefont {Fu}}]{reddy2023fractional}%
  \BibitemOpen
  \bibfield  {author} {\bibinfo {author} {\bibfnamefont {A.~P.}\ \bibnamefont {Reddy}}, \bibinfo {author} {\bibfnamefont {F.}~\bibnamefont {Alsallom}}, \bibinfo {author} {\bibfnamefont {Y.}~\bibnamefont {Zhang}}, \bibinfo {author} {\bibfnamefont {T.}~\bibnamefont {Devakul}},\ and\ \bibinfo {author} {\bibfnamefont {L.}~\bibnamefont {Fu}},\ }\href {https://doi.org/10.1103/PhysRevB.108.085117} {\bibfield  {journal} {\bibinfo  {journal} {Phys. Rev. B}\ }\textbf {\bibinfo {volume} {108}},\ \bibinfo {pages} {085117} (\bibinfo {year} {2023})}\BibitemShut {NoStop}%
\bibitem [{\citenamefont {Paul}\ \emph {et~al.}(2025)\citenamefont {Paul}, \citenamefont {Abouelkomsan}, \citenamefont {Reddy},\ and\ \citenamefont {Fu}}]{paul2025shining}%
  \BibitemOpen
  \bibfield  {author} {\bibinfo {author} {\bibfnamefont {N.}~\bibnamefont {Paul}}, \bibinfo {author} {\bibfnamefont {A.}~\bibnamefont {Abouelkomsan}}, \bibinfo {author} {\bibfnamefont {A.}~\bibnamefont {Reddy}},\ and\ \bibinfo {author} {\bibfnamefont {L.}~\bibnamefont {Fu}},\ }\href@noop {} {\bibfield  {journal} {\bibinfo  {journal} {arXiv preprint arXiv:2502.17569}\ } (\bibinfo {year} {2025})}\BibitemShut {NoStop}%
\bibitem [{\citenamefont {Ledwith}\ \emph {et~al.}(2020{\natexlab{b}})\citenamefont {Ledwith}, \citenamefont {Tarnopolsky}, \citenamefont {Khalaf},\ and\ \citenamefont {Vishwanath}}]{ledwith2020fractional}%
  \BibitemOpen
  \bibfield  {author} {\bibinfo {author} {\bibfnamefont {P.~J.}\ \bibnamefont {Ledwith}}, \bibinfo {author} {\bibfnamefont {G.}~\bibnamefont {Tarnopolsky}}, \bibinfo {author} {\bibfnamefont {E.}~\bibnamefont {Khalaf}},\ and\ \bibinfo {author} {\bibfnamefont {A.}~\bibnamefont {Vishwanath}},\ }\href {https://doi.org/10.1103/PhysRevResearch.2.023237} {\bibfield  {journal} {\bibinfo  {journal} {Phys. Rev. Research}\ }\textbf {\bibinfo {volume} {2}},\ \bibinfo {pages} {023237} (\bibinfo {year} {2020}{\natexlab{b}})}\BibitemShut {NoStop}%
\bibitem [{\citenamefont {{Morales-Dur{\'a}n}}\ \emph {et~al.}(2024)\citenamefont {{Morales-Dur{\'a}n}}, \citenamefont {Wei}, \citenamefont {Shi},\ and\ \citenamefont {MacDonald}}]{morales-duran2024magic}%
  \BibitemOpen
  \bibfield  {author} {\bibinfo {author} {\bibfnamefont {N.}~\bibnamefont {{Morales-Dur{\'a}n}}}, \bibinfo {author} {\bibfnamefont {N.}~\bibnamefont {Wei}}, \bibinfo {author} {\bibfnamefont {J.}~\bibnamefont {Shi}},\ and\ \bibinfo {author} {\bibfnamefont {A.~H.}\ \bibnamefont {MacDonald}},\ }\href {https://doi.org/10.1103/PhysRevLett.132.096602} {\bibfield  {journal} {\bibinfo  {journal} {Phys. Rev. Lett.}\ }\textbf {\bibinfo {volume} {132}},\ \bibinfo {pages} {096602} (\bibinfo {year} {2024})}\BibitemShut {NoStop}%
\bibitem [{\citenamefont {Shi}\ \emph {et~al.}(2024)\citenamefont {Shi}, \citenamefont {Morales-Dur\'an}, \citenamefont {Khalaf},\ and\ \citenamefont {MacDonald}}]{TMDAharonocCasher}%
  \BibitemOpen
  \bibfield  {author} {\bibinfo {author} {\bibfnamefont {J.}~\bibnamefont {Shi}}, \bibinfo {author} {\bibfnamefont {N.}~\bibnamefont {Morales-Dur\'an}}, \bibinfo {author} {\bibfnamefont {E.}~\bibnamefont {Khalaf}},\ and\ \bibinfo {author} {\bibfnamefont {A.~H.}\ \bibnamefont {MacDonald}},\ }\href {https://doi.org/10.1103/PhysRevB.110.035130} {\bibfield  {journal} {\bibinfo  {journal} {Phys. Rev. B}\ }\textbf {\bibinfo {volume} {110}},\ \bibinfo {pages} {035130} (\bibinfo {year} {2024})}\BibitemShut {NoStop}%
\bibitem [{\citenamefont {Dong}\ \emph {et~al.}(2023)\citenamefont {Dong}, \citenamefont {Wang}, \citenamefont {Ledwith}, \citenamefont {Vishwanath},\ and\ \citenamefont {Parker}}]{dong2023composite}%
  \BibitemOpen
  \bibfield  {author} {\bibinfo {author} {\bibfnamefont {J.}~\bibnamefont {Dong}}, \bibinfo {author} {\bibfnamefont {J.}~\bibnamefont {Wang}}, \bibinfo {author} {\bibfnamefont {P.~J.}\ \bibnamefont {Ledwith}}, \bibinfo {author} {\bibfnamefont {A.}~\bibnamefont {Vishwanath}},\ and\ \bibinfo {author} {\bibfnamefont {D.~E.}\ \bibnamefont {Parker}},\ }\href {https://doi.org/10.1103/PhysRevLett.131.136502} {\bibfield  {journal} {\bibinfo  {journal} {Phys. Rev. Lett.}\ }\textbf {\bibinfo {volume} {131}},\ \bibinfo {pages} {136502} (\bibinfo {year} {2023})}\BibitemShut {NoStop}%
\bibitem [{\citenamefont {Mera}\ and\ \citenamefont {Ozawa}(2021{\natexlab{b}})}]{meraKahlerGeometryChern2021}%
  \BibitemOpen
  \bibfield  {author} {\bibinfo {author} {\bibfnamefont {B.}~\bibnamefont {Mera}}\ and\ \bibinfo {author} {\bibfnamefont {T.}~\bibnamefont {Ozawa}},\ }\href {https://doi.org/10.1103/PhysRevB.104.045104} {\bibfield  {journal} {\bibinfo  {journal} {Phys. Rev. B}\ }\textbf {\bibinfo {volume} {104}},\ \bibinfo {pages} {045104} (\bibinfo {year} {2021}{\natexlab{b}})}\BibitemShut {NoStop}%
\bibitem [{\citenamefont {Ledwith}\ \emph {et~al.}(2022{\natexlab{c}})\citenamefont {Ledwith}, \citenamefont {Vishwanath},\ and\ \citenamefont {Khalaf}}]{ledwith2022family}%
  \BibitemOpen
  \bibfield  {author} {\bibinfo {author} {\bibfnamefont {P.~J.}\ \bibnamefont {Ledwith}}, \bibinfo {author} {\bibfnamefont {A.}~\bibnamefont {Vishwanath}},\ and\ \bibinfo {author} {\bibfnamefont {E.}~\bibnamefont {Khalaf}},\ }\href {https://doi.org/10.1103/PhysRevLett.128.176404} {\bibfield  {journal} {\bibinfo  {journal} {Phys. Rev. Lett.}\ }\textbf {\bibinfo {volume} {128}},\ \bibinfo {pages} {176404} (\bibinfo {year} {2022}{\natexlab{c}})}\BibitemShut {NoStop}%
\bibitem [{\citenamefont {Ledwith}\ \emph {et~al.}(2023{\natexlab{a}})\citenamefont {Ledwith}, \citenamefont {Vishwanath},\ and\ \citenamefont {Parker}}]{ledwith_vortexability_2023}%
  \BibitemOpen
  \bibfield  {author} {\bibinfo {author} {\bibfnamefont {P.~J.}\ \bibnamefont {Ledwith}}, \bibinfo {author} {\bibfnamefont {A.}~\bibnamefont {Vishwanath}},\ and\ \bibinfo {author} {\bibfnamefont {D.~E.}\ \bibnamefont {Parker}},\ }\href {https://doi.org/10.1103/PhysRevB.108.205144} {\bibfield  {journal} {\bibinfo  {journal} {Physical Review B}\ }\textbf {\bibinfo {volume} {108}},\ \bibinfo {pages} {205144} (\bibinfo {year} {2023}{\natexlab{a}})},\ \bibinfo {note} {arXiv:2209.15023 [cond-mat]}\BibitemShut {NoStop}%
\bibitem [{\citenamefont {Aharonov}\ and\ \citenamefont {Casher}(1979)}]{aharonov1979ground}%
  \BibitemOpen
  \bibfield  {author} {\bibinfo {author} {\bibfnamefont {Y.}~\bibnamefont {Aharonov}}\ and\ \bibinfo {author} {\bibfnamefont {A.}~\bibnamefont {Casher}},\ }\href {https://doi.org/10.1103/PhysRevA.19.2461} {\bibfield  {journal} {\bibinfo  {journal} {Phys. Rev. A}\ }\textbf {\bibinfo {volume} {19}},\ \bibinfo {pages} {2461} (\bibinfo {year} {1979})}\BibitemShut {NoStop}%
\bibitem [{\citenamefont {Ledwith}\ \emph {et~al.}(2023{\natexlab{b}})\citenamefont {Ledwith}, \citenamefont {Vishwanath},\ and\ \citenamefont {Parker}}]{ledwith2023vortexabilitya}%
  \BibitemOpen
  \bibfield  {author} {\bibinfo {author} {\bibfnamefont {P.~J.}\ \bibnamefont {Ledwith}}, \bibinfo {author} {\bibfnamefont {A.}~\bibnamefont {Vishwanath}},\ and\ \bibinfo {author} {\bibfnamefont {D.~E.}\ \bibnamefont {Parker}},\ }\href {https://doi.org/10.1103/PhysRevB.108.205144} {\bibfield  {journal} {\bibinfo  {journal} {Phys. Rev. B}\ }\textbf {\bibinfo {volume} {108}},\ \bibinfo {pages} {205144} (\bibinfo {year} {2023}{\natexlab{b}})}\BibitemShut {NoStop}%
\bibitem [{\citenamefont {Haldane}(1983)}]{Haldane_1983}%
  \BibitemOpen
  \bibfield  {author} {\bibinfo {author} {\bibfnamefont {F.~D.~M.}\ \bibnamefont {Haldane}},\ }\href {https://doi.org/10.1103/PhysRevLett.51.605} {\bibfield  {journal} {\bibinfo  {journal} {Phys. Rev. Lett.}\ }\textbf {\bibinfo {volume} {51}},\ \bibinfo {pages} {605} (\bibinfo {year} {1983})}\BibitemShut {NoStop}%
\bibitem [{fDi()}]{fDirac}%
  \BibitemOpen
  \href@noop {} {\bibinfo {title} {{Alternatively, we can consider the LLL for a Dirac Hamiltonian which leads to the same zero mode wavefunctions}}}\BibitemShut {NoStop}%
\bibitem [{\citenamefont {Trugman}\ and\ \citenamefont {Kivelson}(1985)}]{trugman-kivelson}%
  \BibitemOpen
  \bibfield  {author} {\bibinfo {author} {\bibfnamefont {S.~A.}\ \bibnamefont {Trugman}}\ and\ \bibinfo {author} {\bibfnamefont {S.}~\bibnamefont {Kivelson}},\ }\href {https://doi.org/10.1103/PhysRevB.31.5280} {\bibfield  {journal} {\bibinfo  {journal} {Phys. Rev. B}\ }\textbf {\bibinfo {volume} {31}},\ \bibinfo {pages} {5280} (\bibinfo {year} {1985})}\BibitemShut {NoStop}%
\bibitem [{\citenamefont {Bernevig}\ and\ \citenamefont {Haldane}(2008)}]{Bernevig_2008}%
  \BibitemOpen
  \bibfield  {author} {\bibinfo {author} {\bibfnamefont {B.~A.}\ \bibnamefont {Bernevig}}\ and\ \bibinfo {author} {\bibfnamefont {F.~D.~M.}\ \bibnamefont {Haldane}},\ }\href {https://doi.org/10.1103/PhysRevLett.100.246802} {\bibfield  {journal} {\bibinfo  {journal} {Phys. Rev. Lett.}\ }\textbf {\bibinfo {volume} {100}},\ \bibinfo {pages} {246802} (\bibinfo {year} {2008})}\BibitemShut {NoStop}%
\bibitem [{\citenamefont {Yang}\ \emph {et~al.}(2012)\citenamefont {Yang}, \citenamefont {Hu}, \citenamefont {Papić},\ and\ \citenamefont {Haldane}}]{Yang_2012}%
  \BibitemOpen
  \bibfield  {author} {\bibinfo {author} {\bibfnamefont {B.}~\bibnamefont {Yang}}, \bibinfo {author} {\bibfnamefont {Z.-X.}\ \bibnamefont {Hu}}, \bibinfo {author} {\bibfnamefont {Z.}~\bibnamefont {Papić}},\ and\ \bibinfo {author} {\bibfnamefont {F.~D.~M.}\ \bibnamefont {Haldane}},\ }\bibfield  {journal} {\bibinfo  {journal} {Physical Review Letters}\ }\textbf {\bibinfo {volume} {108}},\ \href {https://doi.org/10.1103/physrevlett.108.256807} {10.1103/physrevlett.108.256807} (\bibinfo {year} {2012})\BibitemShut {NoStop}%
\bibitem [{Note1()}]{Note1}%
  \BibitemOpen
  \bibinfo {note} {Since we are working in the space of zero modes of $\protect \hat V_1$, we can use $\protect \hat V$ and $\Delta \protect \hat V$ interchangeably}\BibitemShut {NoStop}%
\bibitem [{fWa()}]{fWang}%
  \BibitemOpen
  \href@noop {} {\bibinfo {title} {{We note that ref.~\cite{Wang_2021} has already discussed the need to evaluate $N$-body operator in the presence of non-uniform magnetic field. To address this, they employed an approximation which is equivalent to replacing the operator $\hat T_R$ by the identity in both numerator and denominator. This amounts to replacing $V(z_i,z_j)$ with $V(z_i,z_j) \Gamma(z_i)^2 \Gamma(z_j)^2$ and evaluating the remaining expectation values assuming uniform field. This is in general an uncontrolled approximation. Our choice of the non-uniform field, on the other hand, allows an \emph{exact} mapping regardless of the degree of non-uniformity.}}}\BibitemShut {Stop}%
\bibitem [{\citenamefont {Girvin}(1984)}]{girvin1984anomalous}%
  \BibitemOpen
  \bibfield  {author} {\bibinfo {author} {\bibfnamefont {S.}~\bibnamefont {Girvin}},\ }\href@noop {} {\bibfield  {journal} {\bibinfo  {journal} {Physical Review B}\ }\textbf {\bibinfo {volume} {30}},\ \bibinfo {pages} {558} (\bibinfo {year} {1984})}\BibitemShut {NoStop}%
\bibitem [{\citenamefont {Fulsebakke}\ \emph {et~al.}(2023)\citenamefont {Fulsebakke}, \citenamefont {Fremling}, \citenamefont {Moran},\ and\ \citenamefont {Slingerland}}]{fulsebakke2023parametrization}%
  \BibitemOpen
  \bibfield  {author} {\bibinfo {author} {\bibfnamefont {J.}~\bibnamefont {Fulsebakke}}, \bibinfo {author} {\bibfnamefont {M.}~\bibnamefont {Fremling}}, \bibinfo {author} {\bibfnamefont {N.}~\bibnamefont {Moran}},\ and\ \bibinfo {author} {\bibfnamefont {J.~K.}\ \bibnamefont {Slingerland}},\ }\href@noop {} {\bibfield  {journal} {\bibinfo  {journal} {SciPost Physics}\ }\textbf {\bibinfo {volume} {14}},\ \bibinfo {pages} {149} (\bibinfo {year} {2023})}\BibitemShut {NoStop}%
\bibitem [{Note2()}]{Note2}%
  \BibitemOpen
  \bibinfo {note} {We choose the negative sign since we anticipate the function $\delta g_\xi (z_1, z_2)$ to be mostly negative as it corresponds to a dip relative to the asymptotic value}\BibitemShut {NoStop}%
\bibitem [{\citenamefont {Xie}\ \emph {et~al.}(1993)\citenamefont {Xie}, \citenamefont {Sarma},\ and\ \citenamefont {He}}]{xie1993elementary}%
  \BibitemOpen
  \bibfield  {author} {\bibinfo {author} {\bibfnamefont {X.}~\bibnamefont {Xie}}, \bibinfo {author} {\bibfnamefont {S.~D.}\ \bibnamefont {Sarma}},\ and\ \bibinfo {author} {\bibfnamefont {S.}~\bibnamefont {He}},\ }\href@noop {} {\bibfield  {journal} {\bibinfo  {journal} {Physical Review B}\ }\textbf {\bibinfo {volume} {47}},\ \bibinfo {pages} {15942} (\bibinfo {year} {1993})}\BibitemShut {NoStop}%
\bibitem [{\citenamefont {Xu}\ \emph {et~al.}(2025)\citenamefont {Xu}, \citenamefont {Ji}, \citenamefont {Wang}, \citenamefont {Trung},\ and\ \citenamefont {Yang}}]{xu2025dynamicsclustersanyonsfractional}%
  \BibitemOpen
  \bibfield  {author} {\bibinfo {author} {\bibfnamefont {Q.}~\bibnamefont {Xu}}, \bibinfo {author} {\bibfnamefont {G.}~\bibnamefont {Ji}}, \bibinfo {author} {\bibfnamefont {Y.}~\bibnamefont {Wang}}, \bibinfo {author} {\bibfnamefont {H.~Q.}\ \bibnamefont {Trung}},\ and\ \bibinfo {author} {\bibfnamefont {B.}~\bibnamefont {Yang}},\ }\href {https://arxiv.org/abs/2505.20257} {\bibinfo {title} {Dynamics of clusters of anyons in fractional quantum hall fluids}} (\bibinfo {year} {2025}),\ \Eprint {https://arxiv.org/abs/2505.20257} {arXiv:2505.20257 [cond-mat.str-el]} \BibitemShut {NoStop}%
\bibitem [{\citenamefont {Gattu}\ and\ \citenamefont {Jain}(2025)}]{gattu2025molecularanyonsfractionalquantum}%
  \BibitemOpen
  \bibfield  {author} {\bibinfo {author} {\bibfnamefont {M.}~\bibnamefont {Gattu}}\ and\ \bibinfo {author} {\bibfnamefont {J.~K.}\ \bibnamefont {Jain}},\ }\href {https://arxiv.org/abs/2505.22782} {\bibinfo {title} {Molecular anyons in fractional quantum hall effect}} (\bibinfo {year} {2025}),\ \Eprint {https://arxiv.org/abs/2505.22782} {arXiv:2505.22782 [cond-mat.str-el]} \BibitemShut {NoStop}%
\bibitem [{\citenamefont {Murugan}\ \emph {et~al.}(2019)\citenamefont {Murugan}, \citenamefont {Shock},\ and\ \citenamefont {Slayen}}]{murugan_notes_2019}%
  \BibitemOpen
  \bibfield  {author} {\bibinfo {author} {\bibfnamefont {J.}~\bibnamefont {Murugan}}, \bibinfo {author} {\bibfnamefont {J.~P.}\ \bibnamefont {Shock}},\ and\ \bibinfo {author} {\bibfnamefont {R.~P.}\ \bibnamefont {Slayen}},\ }\href {https://doi.org/10.48550/arXiv.1909.08042} {\bibinfo {title} {Notes on the {Squashed} {Sphere} {Lowest} {Landau} {Level}}} (\bibinfo {year} {2019}),\ \bibinfo {note} {arXiv:1909.08042 [hep-th]}\BibitemShut {NoStop}%
\bibitem [{\citenamefont {Iengo}\ and\ \citenamefont {Li}(1994)}]{Iengo_1994}%
  \BibitemOpen
  \bibfield  {author} {\bibinfo {author} {\bibfnamefont {R.}~\bibnamefont {Iengo}}\ and\ \bibinfo {author} {\bibfnamefont {D.}~\bibnamefont {Li}},\ }\href {https://doi.org/10.1016/0550-3213(94)90010-8} {\bibfield  {journal} {\bibinfo  {journal} {Nuclear Physics B}\ }\textbf {\bibinfo {volume} {413}},\ \bibinfo {pages} {735–753} (\bibinfo {year} {1994})}\BibitemShut {NoStop}%
\bibitem [{\citenamefont {Wu}\ and\ \citenamefont {Yang}(1976)}]{Wu_1976}%
  \BibitemOpen
  \bibfield  {author} {\bibinfo {author} {\bibfnamefont {T.~T.}\ \bibnamefont {Wu}}\ and\ \bibinfo {author} {\bibfnamefont {C.~N.}\ \bibnamefont {Yang}},\ }\href {https://doi.org/10.1016/0550-3213(76)90143-7} {\bibfield  {journal} {\bibinfo  {journal} {Nucl. Phys. B}\ }\textbf {\bibinfo {volume} {107}},\ \bibinfo {pages} {365} (\bibinfo {year} {1976})}\BibitemShut {NoStop}%
\bibitem [{\citenamefont {Wu}\ and\ \citenamefont {Yang}(1977)}]{Wu_1977}%
  \BibitemOpen
  \bibfield  {author} {\bibinfo {author} {\bibfnamefont {T.~T.}\ \bibnamefont {Wu}}\ and\ \bibinfo {author} {\bibfnamefont {C.~N.}\ \bibnamefont {Yang}},\ }\href {https://doi.org/10.1103/PhysRevD.16.1018} {\bibfield  {journal} {\bibinfo  {journal} {Phys. Rev. D}\ }\textbf {\bibinfo {volume} {16}},\ \bibinfo {pages} {1018} (\bibinfo {year} {1977})}\BibitemShut {NoStop}%
\end{thebibliography}%

\onecolumngrid
\appendix
\newpage

\section{Lowest Landau level on the sphere}
\subsection{Stereographic projection}

We perform a stereographic mapping from spherical coordinates to complex coordinates $z=x +iy$, $\bar z=x-iy$. The complex coordinate $z$ is given in terms of the azimuthal and polar angles $(\phi,\theta)$ by
\begin{equation}
    z = r e^{i \phi} = \cot \bigg(\frac{\theta}{2} \bigg)e^{i\phi}.
\end{equation}

We also define the Haldane spinor to be
\begin{equation}
    u = \cos\frac{\theta}{2} e^{i\phi,} \quad v = \sin\frac{\theta}{2},
\end{equation}
such that $z = u/v$. These will be useful for writing down many-body wavefunctions.

We can compute how the metric is transformed from the sphere to the complex plane. The line element $ds$ on the sphere is given in terms of $(\phi,\theta)$ by 
\begin{equation}
    d^2s = d^2 \theta + \sin^2\theta d^2\phi.
\end{equation}

Using $d\theta = 2dr/(1+r^2)$, we can easily convert this to
\begin{equation}
    d^2s = \frac{4}{(1+r^2)^2}(d^2r + r^2 d^2\phi) = \frac{4}{(1+r^2)^2}(d^2x + d^2y),
\end{equation}
which coincides with the usual line element $d^2s = d^2r +r^2d^2\phi$ of the plane up to the position-dependent factor. In the usual cartesian coordinates, the metric is therefore given by
\begin{equation}
    g = \begin{pmatrix}
        \frac{4}{(1+r^2)^2} & 0 \\
        0 & \frac{4}{(1+r^2)^2}
    \end{pmatrix}.
    \label{eq:metric-plane}
\end{equation}
The associated volume form is $\sqrt{g} \equiv \sqrt{|\det g|} = \frac{4}{(1+r^2)^2}$. Unless otherwise specified, $g$ will denote the metric in the cartesian coordinates in the following.

We next map the Laplacian on the sphere to the plane. The Laplacian on the sphere is given by
\begin{equation}
    \Delta = \frac{1}{\sin\theta}\partial_\theta \left(\sin \theta \partial_\theta\right) + \frac{1}{\sin^2 \theta} \partial^2_\phi.
\end{equation}
Using $\partial_\phi \to \partial_\phi$ and $\partial_\theta \to \frac{(1+r^2)}{2} \partial_r$, we get
\begin{equation}
    \Delta_g = \frac{1}{\sqrt{g}} \frac{1}{r}\partial_r \left(r \partial_r\right) + \frac{1}{\sqrt{g}} \frac{1}{r^2}\partial_\phi^2 = \frac{1}{\sqrt{g}} \Delta_{\mathrm{plane}},
\end{equation}
where $\Delta_{\mathrm{plane}}$ is the usual Laplacian on the plane:
\begin{equation}
    \Delta_{\mathrm{plane}} = \frac{1}{r} \partial_r \left( r \partial_r\right) + \frac{1}{r^2} \partial_\phi^2 = \partial_x^2 + \partial_y^2
\end{equation}

$\Delta_g$ is in fact the Laplace-Beltrami operator with the metric $g_{ab}$ given by Eq.~\eqref{eq:metric-plane}:
\begin{equation}
    \Delta_g = \frac{1}{\sqrt{g}}\partial_a (\sqrt{g}g^{ab} \partial_b) = \frac{1}{\sqrt{g}} (\partial_x^2 + \partial_y^2).
\end{equation}

Let us define complex derivatives as
\begin{align}
    \partial_z &= \frac{1}{2}(\partial_x - i \partial_y), \quad
         \partial_{\bar{z}} = \frac{1}{2}(\partial_x + i \partial_y)
\end{align}

such that we can write the Laplacian in complex coordinates as
\begin{equation}
    \partial_x^2 + \partial_y^2 = 4\partial_z \partial_{\bar{z}}.
\end{equation}

On the stereographically projected plane, we instead get
\begin{equation}
    \Delta_g = \frac{4}{\sqrt{g}}  \partial_z \partial_{\bar{z}} = (1+|z|^2)^2 \partial_z \partial_{\bar{z}}.
    \label{eq:stereographic_laplacian}
\end{equation}

The integration measure can be written as
\begin{equation}
    \int \sqrt{g} dx dy = \int \frac{4dx dy}{(1 + r^2)^2} = 
    \int \frac{dz d\bar{z}}{2i(1+|z|^2)^2} \equiv \int d\mu(z).
\end{equation}

\subsection{Uniform and non-uniform magnetic fields on the sphere}
In this section, we mainly follow~\cite{murugan_notes_2019}. The sphere is a K\"ahler manifold, since it is equipped with a complex structure, a Riemannian structure, and a symplectic structure. The corresponding K\"ahler potential $K(|z|)$ is a smooth, real-valued function characterized by 

\begin{equation}
    \partial_z\partial_{\bar{z}} K(|z|) = \sqrt{g}.
\end{equation}

For the planar and spherical geometries, the K\"ahler potential is given by 

\begin{equation} 
    K(|z|)=
    \begin{cases}
        |z|^2 \qquad \qquad \qquad \, \, \text{plane}, \\
        4\log(1+|z|^2) \qquad \text{sphere}.
    \end{cases}
\end{equation}

The magnetic field $\bm{B} = B \hat{\bm{\Omega}} = \frac{2s\Phi_0}{4\pi} \hat{\bm{\Omega}}$ in which the particles move is given in terms of the holomorphic and anti-holomorphic components $A_z=A_x+iA_y$ and $A_{\bar{z}}=A_x-iA_y$ of the magnetic vector potential as

\begin{equation}
    B = \frac{2i}{\sqrt{g}}(\partial_{\bar{z}}A_z-\partial_zA_{\bar{z}}) = \frac{4i}{\sqrt{g}} \partial_{\bar{z}} A_z,
    \label{magnetic_field_vector_pot}
\end{equation}
where for the last equality we used the Coulomb gauge $\partial_{\bar{z}}A_z + \partial_z A_{\bar{z}} = 0$.  If we define the \textit{magnetic potential} $Q(|z|)$ as

\begin{gather}
    i\hbar\partial_zQ(|z|)= -2eA_z, \quad -i\hbar\partial_{\bar{z}}Q(|z|) = 2eA_{\bar{z}}
\end{gather}

and combine this with equation \ref{magnetic_field_vector_pot}, we obtain a second order differential equation for the magnetic potential 

\begin{equation}
    \Delta_g Q(|z|)= \frac{4}{\sqrt{g}}\partial_z\partial_{\bar{z}} Q(|z|) = - \frac{2e}{\hbar}B(|z|).
\end{equation}
where $\Delta_g$ is the Laplace-Beltrami operator. For simplicity, in the rest of the paper we will set $e = \hbar = 1$. \\
In the case of homogeneous magnetic flux through the spherical surface, the magnetic potential $Q(|z|)$ can be chosen to be proportional to the Kähler potential as
\begin{equation}
    Q(|z|) = - \frac{s}{2}K(|z|) = -2s \log(1+|z|^2)
    \label{uniform_magpot}
\end{equation}
where $s$ is the magnetic monopole strength.

To make the magnetic field inhomogeneous, we deform the magnetic potential:
\begin{equation}
    Q^R(|z|) = 2 \log (1+|z|^2) -2(s+1) \log(1+|z|^2/R^2).
    \label{magpot}
\end{equation}
For $R=1$, this reduces to the homogeneous magnetic field in equation \ref{uniform_magpot}. The corresponding magnetic field is then given by 
\begin{equation}
    B^R(|z|) = \frac{\Phi_0}{4\pi} \biggl( 2(s+1) \frac{(1+|z|^2)^2}{(1+|z|^2/R^2)^2} - 2 \biggr).
    \label{magfluxdens}
\end{equation} 

To see that this has the same magnetic field as a function of $R$, we note that the difference of the magnetic potential
\begin{equation}
    \Delta Q^R(z) = Q^R(z) - Q(z) = 2(s+1)\log\frac{1+|z|^2}{1+|z|^2/R^2},
\end{equation}
has no zero or singularity. In fact, this allows for a wide choice of non-uniform magnetic fields:
\begin{equation}
    Q^R_a(|z|) = 2 (a + 1) \log (1+|z|^2) -2(s+a+1) \log(1+|z|^2/R^2).
\end{equation}
We focus on the $a=0$ case in our paper, as it gives us analytical handle over anyon dispersion.

\subsection{LLL wave functions on the sphere}
The Pauli Hamiltonian for spin polarized electrons on a Riemann surface is given in terms of the holomorphic and anti-holomorphic components of the kinetic momentum $\pi_z=-i\hbar\partial_z-eA_z, \, \pi_{\bar z}=-i\hbar\partial_{\bar z}-eA_{\bar z}$ by \cite{Iengo_1994}

\begin{equation}
    H=\frac{1}{2m} \biggl( \frac{1}{\sqrt{g}} \pi_z \pi_{\bar z} + \frac{2-g_s}{8}e\hbar B \biggr),
\end{equation}
where $g_s$ is the Lande g-factor which is approximately $2$ for spin-$1/2$ particles. Using the magnetic potential, we can rewrite the Hamiltonian

\begin{equation}
    H=-\frac{2}{\sqrt{g}} \biggl( \partial_z - \frac{1}{2}\partial_z Q^R(|z|) \biggr) \biggl( \partial_{\bar z} + \frac{1}{2} \partial_{\bar z} Q^R(|z|) \biggr) + \frac{2-g_s}{4} B.
\end{equation}

It is now natural to introduce the transformation $\tilde H = e^{\frac{1}{2}Q^R} H e^{-\frac{s}{2}Q^R}$ such that the transformed Hamiltonian reads

\begin{equation}
    \tilde H = -\frac{2}{\sqrt{g}} \biggl( \partial_z - \partial_z Q^R(|z|) \biggr) \partial_{\bar z} + \frac{2-g_s}{4} B.
    \label{Kähler_Hami}
\end{equation}

Clearly, any holomorphic wave function is a zero mode of the first term of the Hamiltonian $\tilde H$. For generic $g_s$, the second term, which is inhomogeneous for $R\neq 1$, spoils the solvability of the Hamiltonian. If we set $g_s=2$, however, these modes are degenerate and constitute the LLL of the free fermion. For $R=1$, the simplest choice of basis for the LLL is then given by \cite{Haldane_1983}
\begin{equation}
    \phi_m(z, \bar z) = \alpha_{s,m}z^{s+m} e^{Q(|z|)/2},
\end{equation}
where $z^{s+m}$ is the holomorphic part satisfying $\partial_{\bar z}z^{s+m}=0$. The range of angular momentum is limited to $m = -s, \ldots,s$ as the wavefunction becomes singular or unnormalizable otherwise.

In a non-uniform field, the LLL eigenstates are given by
\begin{equation}
    \phi^R_{s,m}(z,\bar z) = e^{-\frac{1}{2}\beta_m} \alpha_{s, m} z^{s+m} e^{\frac{1}{2}Q^R(|z|)} 
    \label{eq:nonuniwave}
\end{equation}
with $Q^R(|z|)$ parameterized as in equation \ref{magpot}. The degeneracy remains $N_\phi+1$ for the LLL basis, with $m=-s,-s+1\dots,s$. The wave functions are normalizable with respect to the inner product 
\begin{align}
    \int \frac{4 dx dy}{(1 + |z|^2)^2}
    \overline{\phi_{s,m}^R(z, \bar{z})}
    \phi_{s,m}^R(z, \bar{z}).
\end{align}

For our choice of magnetic potential $Q_R(|z|)$ in equation \ref{magpot}, this reduces to a simple form which can be calculated as follows
\begin{align}
    e^{\beta_m} &= |\alpha_{s, m}|^2\int \frac{4 dx dy}{(1 + |z|^2)^2} \frac{|z|^{2(s+m)}(1+|z|^2)^2}{(1+|z|^2/R^2)^{2(s+1)}} \\
    &= |\alpha_{s, m}|^2\int 4dxdy\frac{|z|^{2(s+m)}}{(1+|z|^2/R^2)^{(2s+1)}} \\
    &= R^{2(s+m+1)} |\alpha_{s, m}|^2\int 4dxdy\frac{|z|^{2(s+m)}}{(1+|z|^2)^{2(s+1)}} \\
    &= R^{2(s+m+1)},
\end{align} 
where we used $|\alpha_{s, m}|^{-2} = \int 4dxdy\frac{|z|^{2(s+m)}}{(1+|z|^2)^{2(s+1)}}$.

\section{Interacting Hamiltonian}
\subsection{Interactions on the sphere}
We consider spherically symmetric interactions, such that the sole source of rotational symmetry breaking arises from the single-particle part of the Hamiltonian. To achieve this, we demand that the interaction only depends on the chord distance:
\begin{equation}
    d(\Omega_1, \Omega_2) = 2 \sin\left(\frac{\Omega_{12}}{2}\right),
\end{equation}
where $\Omega_{12}$ is the angle between $\Omega_1$ and $\Omega_2$. The Coulomb interaction is given by
\begin{equation}
    V_{\mathrm{Coulomb}}(\Omega_1, \Omega_2) = \frac{1}{d(\Omega_1, \Omega_2)}.
\end{equation}
The $n$-th pseudopotential, which can be used to expand short-range interaction~\cite{trugman-kivelson}, is defined by
\begin{equation}
    V^n(\Omega_1, \Omega_2) = \frac{1}{(n!)^2}(\Delta)^{n} \delta(\Omega_1, \Omega_2),
\end{equation}
where $\Delta$ is the standard Laplacian on the sphere. For the spin-polarized fermionic problem of interest, only odd $n$ contribute.

To map this to the plane, we first note that the distance function can be written as
\begin{equation}
    d(z_1, z_2) = \frac{2|z_1 - z_2|}{\sqrt{1 + |z_1|^2} \sqrt{1 + |z_2|^2}}.
    \label{eq:stereographic-distance}
\end{equation}
The form of the Coulomb interaction immediately follows.
The pseudopotential can be written as
\begin{equation}
    \frac{1}{(n!)^2}\Delta_g^n \left(\frac{(1+|z|^2)^2}{4}\delta(x_1 - x_2)\delta(y_1 - y_2)\right),
\end{equation}
where we used the mapping
\begin{equation}
    \delta_g(z_1, z_2) = \frac{(1+|z|^2)^2}{4}\delta(x_1 - x_2)\delta(y_1 - y_2).
\end{equation}
The correctness of this map can be confirmed by noting
\begin{equation}
    \int d\mu(z) \delta_g(z, z_0) f(z) = f(z_0),
\end{equation}
as expected for a delta function.

\subsection{Second quantized form of the interaction}
For the numerical ED, we evaluate $2$-body scattering elements between fermions in the LLL. Let us start by writing the most general form of a $2$-body interaction in the angular momentum basis in second quantization

\begin{equation}
    \hat{V} = \sum_{\ell=1}^{4s-1}\sum_{m,m'=-s}^{s} V_{\ell;mm'} c^\dagger_m c^\dagger_{\ell-m}c_{m'}c_{\ell-m'},
\end{equation}
where $\ell$ is the total $L_z$ angular momentum of the particle pair and $m$ and $m'$ are their relative angular momenta. The interaction matrix element $V_{\ell;mm'}$ can be calculated from the following integral
\begin{align}
    V_{\ell;mm'} = \int d\mu(z_1) d\mu(z_2) \phi_{s,m}^{R,*}(z_1,\bar z_1) \phi_{s,\ell-m}^{R, *}(z_2,\bar z_2) V(z_1, z_2) \phi^R_{s,\ell-m'}(z_2,\bar z_2) \phi^R_{s,m'}(z_1,\bar z_1) .
    \label{interaction_integral}
\end{align}

For spherically symmetric interactions, we can expand $V(z_1-z_2)$ in terms of the complete and orthonormal set of spherical harmonics $\Y_{\ell}^m$ as 

\begin{align}
    V(z_1, z_2) &= \sum_{\ell=0}^\infty\sum_{m=-\ell}^\ell V_\ell \bar \Y_{\ell}^{m}(z_1,\bar z_1) \Y_{\ell}^{m}(z_2,\bar z_2).
    \label{spherical_expansion}
\end{align} 

The spherical harmonics $\Y_{\ell,m}$ are joint eigenfunctions of the angular momentum operators $\bm{L}^2$ and $L_z$ with eigenvalues $-\ell(\ell+1)$ and $m$ respectively.

The interaction matrix element given in equation \ref{interaction_integral} therefore separates into two independent integrals and can be written as
\begin{equation}
    V_{\ell;mm'} = \sum_{\ell'=0}^{2s}\sum_{m''=-\ell}^\ell V_{\ell'} \bar \lambda_{\ell,m;R}^{\ell',m''} \lambda_{\ell,m';R}^{\ell',m''},
\end{equation}
where the form factors $\lambda_{\ell,m';R}^{\ell',m''}$ are given by 
\begin{align}
    \lambda_{\ell, m'}^{\ell',m'';R} &=
    \int d\mu(z) \phi^R_{s,m'}(z,\bar z)\phi_{s,\ell-m'}^R  (z, \bar z)Y_{\ell'}^{m''}(z, \bar z)
\end{align}

In the spherically symmetric case, these form factors have been evaluated \cite{Wu_1976,Wu_1977} 
\begin{align}
    \lambda_{\ell, m'}^{\ell',m''} &=
    \int d\mu(z) \phi_{s,m'}(z,\bar z)\phi_{s,\ell-m'}(z, \bar z)Y_{\ell'}^{m''}(z, \bar z) \\
    &= (-1)^{2s+\ell} (2s+1)\bigg(\frac{(2\ell+1)}{4\pi} \bigg)^{1/2}
    \begin{pmatrix}
        s & s & \ell' \\
        m' & \ell-m' & -m''
    \end{pmatrix}
    \begin{pmatrix}
        s & s & \ell' \\
        s & s & 0
    \end{pmatrix},
\end{align}
where the matrices are $3$-j symbols that are proportional to the Clebsch-Gordan coefficients. In order for the $3$-j symbols to be non-zero, the triangle relation $0<\ell'\leq2s$ has to be fulfilled. Furthermore, momentum conservation guarantees that $\ell-m'=0$. In the case of non-uniform magnetic field, we calculate the form factors numerically. 

Let us finally compute $V_\ell$ for Coulomb and pseudopotential interactions. It is convenient to compute $V_\ell$ in the spherical basis, and then map things to the plane.
The long-ranged Coulomb interaction can be expanded as

\begin{equation}
    \frac{1}{d(\bm\Omega_1, \bm \Omega_2)} = \frac{1}{\sqrt{2-2\cos\theta_{12}}} = \sum_{\ell=0}^\infty \P_\ell(\cos\theta_{12}),
\end{equation}
where $\theta_{12}$ is the angle between the particles on the sphere. Using the spherical harmonic addition theorem 
\begin{equation}
    \P_\ell(\cos\theta_{12}) = \frac{4\pi}{2\ell+1}\sum_{m=-\ell}^\ell \bar \Y_{\ell}^m(z_1,\bar z_1) \Y_{\ell}^m(z_2,\bar z_2),
\end{equation}
we get 
\begin{equation}
    V_{\ell, \mathrm{Coulomb}} = \frac{4\pi}{2\ell + 1}.
\end{equation}
The pseudopotential, on the other hand, can be obtained by expanding the delta function and acting with the Laplacian on it:
\begin{align}
    V^n &= \frac{1}{(n!)^2}\sum_{n=0}^\infty \sum_{\ell,m} \Delta^{n} \bar \Y_{\ell}^m(\Omega_1) \Y_{\ell}^m(\Omega_2) \\
    &= \frac{1}{(n!)^2}\sum_{n=0}^\infty \sum_{\ell,m} (-\ell(\ell+1))^n \bar \Y_{\ell}^m(\Omega_1) \Y_{\ell}^m(\Omega_2),
\end{align}
from which we can read off
\begin{equation}
    V^n_\ell = \frac{1}{(n!)^2}(-\ell(\ell+1))^n.
\end{equation}

\subsection{Transforming the Hamiltonian}
The treatment above was quite general, and it applies to any choice of inhomogeneous magnetic field, not just the one we used in the main text. Our choice for $Q^R$, however, allows significant simplifications.

To see this, let us evaluate the second quantized Hamiltonian slightly differently. To do so, let us first rewrite the density-density interaction:

\begin{align}
    :\hat\rho_{z_1}^R \hat\rho_{z_2}^R:
    &= c^{\dagger}_{R;z_1} c^{\dagger}_{R;_2} c_{R;z_2} c_{R;z_1}
    \\
    &=\sum_{m_1, m_2, m_3, m_4}
    \overline{\phi^R_{m_1}(z_1)}
     \overline{\phi^R_{m_2}(z_2)}
     \phi^R_{m_3}(z_2)
     \phi^R_{m_4}(z_1)
    c^\dagger_{m_1} c^\dagger_{m_2} c_{m_3} c_{m_4},
\end{align}
where we used the projected position operator $c_{R;z} = \sum_m \phi^R_m(z) c_m$. Note that $c_m$ does not carry $R$ labels, since their commutation relations do not depend on $R$.

Integrating this against the interaction, we get
\begin{align}
    &\int d\mu(z_1) d\mu(z_2)
    V(z_1, z_2) :\hat\rho_{z_1}^R \hat\rho_{z_2}^R:
    \\
    &= \int d\mu(z_1) d\mu(z_2)
    V(z_1, z_2) \sum_{m_1, m_2, m_3}
    \overline{\phi^R_{m_1}(z_1)}
     \overline{\phi^R_{m_2}(z_2)}
     \phi^R_{m_3}(z_2)
     \phi^R_{m_1 + m_2 - m_3}(z_1)
    c^\dagger_{m_1} c^\dagger_{m_2} c_{m_3} c_{m_4},
    \\
    &= \int d\mu(z_1) d\mu(z_2)
    V(z_1, z_2) R^{-4} \frac{(1+|z_1|^2)^2}{(1+\frac{|z_1|^2}{R^2})^2}
    \frac{(1+|z_2|^2)^2}{(1+\frac{|z_2|^2}{R^2})^2}
    \sum_{m_1, m_2, m_3}
    \overline{\phi_{m_1}(\frac{z_1}{R})}
     \overline{\phi_{m_2}(\frac{z_2}{R})}
     \phi_{m_3}(\frac{z_2}{R})
     \phi_{m_1 + m_2 - m_3}(\frac{z_1}{R})
    c^\dagger_{m_1} c^\dagger_{m_2} c_{m_3} c_{m_4},
    \\
    &= \int d\mu(z_1) d\mu(z_2)
    V(Rz_1, Rz_2)
    \sum_{m_1, m_2, m_3}
    \overline{\phi_{m_1}(z_1)}
     \overline{\phi_{m_2}(z_2)}
     \phi_{m_3}(z_2)
     \phi_{m_1 + m_2 - m_3}(z_1)
    c^\dagger_{m_1} c^\dagger_{m_2} c_{m_3} c_{m_4}
\end{align}
where we used

\begin{equation}
\phi^R_m(z_1)
= e^{-\frac{\beta_m}{2}}
\frac{\displaystyle \left(1 + \lvert z_1 \rvert^2\right)^{(s+1)}}
     {\displaystyle \left(1 + \frac{\lvert z_1 \rvert^2}{R^2}\right)^{(s+1)}}
\;\phi_m(z_1)
= \alpha_{s,m}\,R^{-(s+m+1)}
  \frac{\displaystyle z_1^{\,s+m}}
       {\displaystyle \left(1 + \frac{\lvert z_1 \rvert^2}{R^2}\right)^{s}}
  \;\frac{\displaystyle \left(1 + \lvert z_1 \rvert^2\right)}
        {\displaystyle \left(1 + \frac{\lvert z_1 \rvert^2}{R^2}\right)}
= R^{-1}\,
  \frac{\displaystyle \left(1 + \lvert z_1 \rvert^2\right)}
       {\displaystyle \left(1 + \frac{\lvert z_1 \rvert^2}{R^2}\right)}
\;\phi_m\!\Bigl(\frac{z_1}{R}\Bigr)\,. 
\end{equation}

Therefore, the spectrum of the interacting Hamiltonian with $V(z_1, z_2)$ and non-uniform field $Q^R$ can be reproduced by the interaction $V(Rz_1, Rz_2)$ with uniform field $Q$.

\section{Mapping to uniform field}
The wavefunctions in the non-uniform magnetic field are given in Eq.~\ref{eq:nonuniwave}. We introduce an invertible, non-unitary transformation in second quantization:

\begin{equation}
    \hat T_R = \hat T_R^\dagger = e^{\frac{1}{2}\sum_m\beta_m^R  \hat c_m^\dagger \hat c_m}, \qquad  \hat T_R \hat c_m \hat T_R^{-1} = e^{-\frac{1}{2}\beta_m^R} \hat c_m.
\end{equation}

The quasihole states $|\Psi_M\rangle$ with total momentum $M$ then transform as
\begin{equation}
    |\Psi^{\rm R}_M \rangle = \hat T_R|\Psi_M \rangle.
\end{equation} 

Now consider a (first-quantized) interaction $V(z_1, z_2)$ and write the anyon dispersion as
\begin{equation}
    \epsilon_m^R = \int d\mu(z_1) d\mu(z_2) V(z_1,z_2) g_M^R(z_1,z_2)
\end{equation}
where $d\mu(z) = \sqrt{g(z)}d^2 z = \frac{4}{1+|z|^2} d^2 z$ is the integration measure and $g_m^R(z_1,z_2)$ is the pair correlation function defined as 
\begin{equation}
    g_M^R(z_1,z_2) = \frac{\langle \Psi_M^R|c_{R;z_1}^\dagger c_{R;z_2}^\dagger c_{R;z_2} c_{R;z_1}|\Psi_M^R \rangle}{\langle \Psi_M^R|\Psi_M^R \rangle}
    \label{gMR}
\end{equation}
where we made the $R$ dependence for the electron creation/annihilation operators explicit. These operators can be expanded in terms of angular momentum operators $c_{R;m}$ as
\begin{equation}
    c_{R;z} = \sum_{m=-s}^s \phi_{s,m}^R(z,\bar z) c_{R;m} = e^{\frac{1}{2} \Delta Q^R(|z|)} \hat T_R c_z \hat T_R^{-1}
\end{equation}
Substituting in (\ref{gMR}) yields
\begin{equation}
    g_M^R(z_1,z_2) = e^{\Delta Q^R(|z_1|)}
    e^{\Delta Q^R(|z_2|)}\frac{\langle \Psi_M|c_{z_1}^\dagger c_{z_2}^\dagger T_R^2 c_{z_2} c_{z_1}|\Psi_M \rangle}{\langle \Psi_M|T_R^2|\Psi_M \rangle}.
\end{equation}
This formula works for any choice of  $Q^R$, as long as we can evaluate the action of $T_R$.

\section{Extracting anyon dispersion}
\subsection{Quasihole wavefunctions in angular momentum and coherent state representations}
The quasiholes $|\Psi_m \rangle$ form a representation of the angular momentum operators. Since there are $N+1$ quasiholes states with $m = -\frac{N}{2},\dots,\frac{N}{2}$, this corresponds to the $l = \frac{N}{2}$ representation. This means that we can express the normalized quasihole wavefunctions in terms of the lowest angular momentum state via
\begin{equation}
    |\Psi_m \rangle = \sqrt{\frac{(l-m)!}{2l!(l+m)!}} L_+^{m+l} |\Psi_{-l} \rangle
\end{equation}
The explicit wavefunction for the quasihole with $m = -l$ can be written as
\begin{equation}
    \Psi_{-l}(\{u,v\}) = C \prod_{i<j} (u_i v_j - u_j v_i)^3 \prod_i v_i
\end{equation}
where $C$ is a normalization constant that only depends on $N$. The raising operator takes the form $L_+ = \sum_i u_i \frac{\partial}{\partial v_i}$. We can verify that it commutes with $\prod_{i<j} (u_i v_j - u_j v_i)^3$. Its action on the last part $\prod_i v_i$ takes the form
\begin{equation}
    L^k_+ \prod_i v_i = k! P_k(\{z_i\}) \prod_i v_i
\end{equation}
where $z = u/v$ and $P_k$ are the symmetric polynomials
\begin{equation}
     P_k(z_1,\dots,z_N) = \sum_{r_1 < r_2 <\dots < r_k} z_{r_1} \dots z_{r_k}
\end{equation}
For concreteness, consider $N=3$. Then 
\begin{gather}
\prod_i v_i = v_1 v_2 v_3, \qquad L_+ \prod_i v_i = v_1 v_2 u_3 + v_1 u_2 v_3 + u_1 v_2 v_3 = (z_1 + z_2 + z_3) v_1 v_2 v_3  \\
L_+^2 \prod_i v_i = 2 (u_1 u_2 v_3 + u_1 v_2 u_3 + v_1 u_2 u_3) = 2 (z_1 z_2 + z_1 z_3 + z_2 z_3) v_1 v_2 v_3 \\
L_+^3 \prod_i v_i = 6 u_1 u_2 u_3 = 6 z_1 z_2 z_3 v_1 v_2 v_3
\end{gather}
We can then write the normalized quasihole wavefunctions at arbitrary angular momentum $m$ as
\begin{equation}
    \Psi_m(\{z_i\}) = \sqrt{\frac{(l-m)!}{2l!(l+m)!}} L_+^{m+l} \Psi_{-l} = \sqrt{\frac{(l-m)!(m+l)!}{2l!}}  P_{m+l}(\{z_i\}) \Psi_{-l}(\{z_i\})
    \label{eq:angular_momentum_and_coherent}
\end{equation}
Now instead of consider angular momentum basis, we can also consider the wavefunction for a quasihole localized at point $\xi$ given by
\begin{equation}
    \Psi_\xi(\{z_i\}) = C_\xi \Psi_{-l}(\{z_i\}) \prod_{i} (z_i - \xi) = C_\xi \Psi_{-l}(\{z_i\}) \sum_{m=-l}^{l} P_{l+m}(\{z_i\}) (-\xi)^{l-m} 
\end{equation}
where $C_\xi$ is a normalization constant that generally depends on $\xi$. Using Eq.~\eqref{eq:angular_momentum_and_coherent}, we can relate the normalized second quantized wavefunctions $|\Psi_\xi \rangle$ and $|\psi_m \rangle$
\begin{equation}
    |\Psi_\xi \rangle = C_\xi  \sum_{m=-l}^{l} \binom{2l}{l-m}^{1/2} (-\xi)^{l-m} |\Psi_m \rangle
\end{equation}
which allows us to determine the normalization constant $C_\xi$ via
\begin{equation}
    1 = \langle \Psi_\xi|\Psi_\xi \rangle = |C_\xi|^2 \sum_{m=-l}^{l} \binom{2l}{l-m} |\xi|^{2(l-m)} = |C_\xi|^2 (1 + |\xi|^2)^{2l}, \quad \implies \quad C_\xi = \frac{1}{(1 + |\xi|^2)^l}
\end{equation}
Another way to write this expression is to note that
\begin{equation}
    \xi^{2l} e^{-\xi^{-1} L_+} |\Psi_{-l} \rangle = \sum_{m=-l}^l \frac{(-\xi)^{l-m}}{(m+l)!} L_+^{m+l } |\Psi_{-l} \rangle = \sum_{m=-l}^l (-\xi)^{l-m} \binom{2l}{l+m}^{1/2} |\Psi_{m} \rangle = (1 + |\xi|^2)^l |\Psi_\xi \rangle
\end{equation}
leading to
\begin{equation}
    |\Psi_\xi \rangle = \frac{\xi^{2l}}{(1 + |\xi|^2)^l} e^{-\xi^{-1} L_+} |\Psi_{-l} \rangle
\end{equation}
This is the familiar expression for spin coherent states.

We can also express $|\Psi_m \rangle$ in terms of $|\Psi_\xi \rangle$ through
\begin{equation}
    |\Psi_m \rangle = \sqrt{\frac{(l+m)!}{(2l)! (l-m)!}} \left(-\frac{d}{d\xi}\right)^{l-m} (1 + |\xi|^2)^l |\Psi_\xi \rangle \Big|_{\xi = 0}
\end{equation}
This allows us to relate the expectation value of any operator $\hat O$ in the states $|\Psi_\xi \rangle$ and $|\Psi_m \rangle$ using
\begin{equation}
    \langle \Psi_m| \hat O|\Psi_m \rangle = \frac{(l+m)!}{(2l)! (l-m)!} \left(\frac{d^2}{d\xi d\bar \xi}\right)^{l-m} (1 + |\xi|^2)^{2l} \langle \Psi_\xi| \hat O |\Psi_\xi \rangle \Big|_{\xi = 0}
\end{equation}
If the operator $\hat O$ commutes with $\hat L_z$, we can also write
\begin{equation}
    \langle \Psi_\xi|\hat O|\Psi_\xi \rangle = \frac{1}{(1 + |\xi|^2)^{2l}} \sum_{m=-l}^l \binom{2l}{l-m} |\xi|^{2(l-m)} \langle \Psi_m|\hat O|\Psi_m \rangle
\end{equation}

In particular, by choosing $\hat{O}$ to be the second quantized form of the interaction $V(Rz_1, Rz_2)$, we can get
\begin{equation}
    \epsilon_m^R= \frac{(l+m)!}{(2l)! (l-m)!} \left(\frac{d^2}{d\xi d\bar \xi}\right)^{l-m} (1 + |\xi|^2)^{2l} v_\xi^R \Big|_{\xi = 0}, \qquad v_\xi^R = \frac{1}{(1 + |\xi|^2)^{2l}} \sum_{m=-l}^l \binom{2l}{l-m} |\xi|^{2(l-m)} \epsilon_m^R
    \label{PotentialToSpectrum}
\end{equation}
We will find it more helpful to derive an alternative inversion formula. For this, it is useful to define the variable
\begin{equation}
    y = \frac{1}{1 + |\xi|^2} = \sin^2 \frac{\theta}{2}
\end{equation}
which goes from 0 at the north pole to 1 at the south pole. We can then use Eq.~\ref{PotentialToSpectrum} to write (we will drop the $R$ superscript for simplicity)
\begin{equation}
    v_y = \sum_{m=-l}^l \binom{2l}{l-m} y^{(l+m)} (1 - y)^{(l-m)} \epsilon_m = \sum_{m=-l}^l B^{2l}_{l+m} \epsilon_m = \sum_{m=-l}^l B^{2l}_{l+m}(y) \epsilon_m
\end{equation}
where $B_m^q(y) := \binom{q}{m} y^m (1-y)^{q-m}$ is the Bernstein basis. For a function expressed in the Bernstein basis, we have the inversion formula
\begin{equation}
    \epsilon_{m} = \sum_{k=0}^{m+l} \frac{v_0^{(k)}}{k!} \binom{m+l}{k} \binom{2l}{k}^{-1},
    \label{BersteinInversion}
\end{equation}
where $v_0^{(k)}$ is the $k$-th term in the taylor series expansion of $v_y$. This expression particularly simplifies in the thermodynamic limit where $l$ is taken to be large. It is then useful to also introduce the variable $\kappa = \frac{m}{l}$ with $-1 \leq \kappa \leq 1$. Then we can simplify
\begin{equation}
    \binom{m+l}{k} \binom{2l}{k}^{-1} = \left(\frac{1 + \kappa}{2} \right)^{k} + O(1/l)
\end{equation}
which leads to the expression for the dispersion
\begin{equation}
    \epsilon^R_\kappa = v^R_{y = \frac{1 + \kappa}{2}} = v^R_{\xi = \sqrt{\frac{1 - \kappa}{1 + \kappa}}}
    \label{InversionThermodynamic}
\end{equation}
where we have restored the $R$-dependence.

\subsection{Laughlin quasihole pair correlation functions}
It follows from Eqs.~\ref{PotentialToSpectrum},\ref{BersteinInversion}, \ref{InversionThermodynamic} that it suffices to compute the anyon potential energy $v_\xi^R$ to determine the anyon dispersion. The latter is expression in terms of the pair correlation function for a Laughlin quasihole at $\xi$ for uniform field
\begin{equation}
    v^R_\xi = \int d\mu(z_1) d\mu(z_2) V(R z_1, R z_2) g_\xi(z_1,z_2), \qquad g_\xi(z_1, z_2) = \langle \psi_\xi|f_{z_1}^\dagger f_{z_2}^\dagger f_{z_2} f_{z_1}|\Psi_\xi \rangle, 
    \label{PotentialPairCorrelation}
\end{equation}
We will find it easier to write the expression for $g_\xi(z_1,z_2)$ using the first-quantized wavefunctions. First, we define
\begin{equation}
    \gamma(z_1,z_2) := \frac{2(z_1 - z_2)}{\sqrt{1 + |z_1|^2}\sqrt{1 + |z_2|^2}} =
    \frac{2(u_1/v_1 - u_2/v_2)}{\sqrt{1 + |u_1/v_1|^2}\sqrt{1 + |u_2/v_2|^2}}
    = 2(u_1 v_2 - u_2 v_1)
\end{equation}
$\gamma(z_1,z_2)$ is a complex function whose absolute value is the arc length defined earlier in Eq.~\ref{eq:stereographic-distance}, $|\gamma(z_1,z_2)| = d(z_1,z_2)$. The Laughlin state is given by 
\begin{equation}
    \Psi(\{z_1,\dots,z_n\}) = \prod_{i<j} \gamma(z_i,z_j)^3.
\end{equation}
Likewise the Laughlin QHs are
\begin{equation}
    \Psi_\xi(\{z_1,\dots,z_n\}) = \prod_i \gamma(z_i, \xi) \prod_{i<j} \gamma(z_i, z_j)^3
\end{equation}
which implies that we can write
\begin{equation}
    g_\xi(z_1, z_2) = d(z_1,z_2)^6 f_\xi(z_1,z_2)
\end{equation}
where 
\begin{equation}
    f_\xi(z_1,z_2) = d(z_1,\xi)^2 d(z_2,\xi)^2  \int \prod_{i=3}^N d\mu(z_i) d(z_1,z_i)^6 d(z_2,z_i)^6 d(z_i,\xi)^2 \prod_{2<i<j} d(z_i, z_j)^6
\end{equation}
While $g_\xi(z_1,z_2)$ is a complicated functions of $\xi$, $z_1$, $z_2$, it is easy to see it is invariant under simultaneously applying any isometry for $z_1$, $z_2$, and $\xi$ since both the distance function $d$ and the measure $\mu$ are invariant under isometries. This means that $g_\xi(z_1, z_2, \xi)$ only depends on the distances between the three points $z_1$, $z_2$, and $\xi$:
\begin{equation}
    g_\xi(z_1, z_2) = g(d(z_1,z_2), d(z_1, \xi), d(z_2, \xi))
\end{equation}
This allows us to express the pair correlation function for a hole at an arbitrary point $\xi$ in terms of the pair correlation function for a hole at the noth or south pole. For simplicity, we choose the south pole $\xi = 0$ and define the m\"obius transformation $w_\xi(z) = \frac{z - \xi}{1 + \bar \xi z}$ which preserves the metric and distance and satisfies $w_\xi(\xi) = 0$ leading to
\begin{equation}
    g_\xi(z_1, z_2) = g_{\rm SP}(w_\xi(z_1), w_\xi(z_2))
    \label{PairCorrelationSouthPole}
\end{equation}
where $g_{\rm NP}(w_1, w_2)$ is the pair correlation function for a hole at the south pole. Note that this does not only depend on the distance between $w_1$ and $w_2$ but also their distance to the south pole. 

\subsection{Dispersion in non-uniform field}
Eq.~\ref{PairCorrelationSouthPole} and \ref{PotentialPairCorrelation} allow us to compute the dispersion for any given interaction from the knowledge of the pair correlation function for a single quasihole localized at the south pole. We do this by moving the dependence of $\xi$ in (\ref{PotentialPairCorrelation}) completely to the interaction
\begin{equation}
    v^R_\xi = \int d\mu(w_1) d\mu(w_2) V(R z_\xi(w_1), R z_\xi(w_2)) g_{\rm SP}(w_1,w_2)
    \label{PotentialSouthPole}
\end{equation}
where $z_\xi(w) := \frac{w + \xi}{1 - \bar \xi w}$ satisfies $z_\xi(w_\xi(z)) = z$. As a consistency check, we can verify that for $R = 1$ and for $V$ invariant under isometries, we get $V(z_\xi(w_1), z_\xi(w_2)) = V(w_1, w_2)$ and the integral is independent on $\xi$ as expected. We can further simplify this expression by assuming the interaction is only a function of the distance $d(z_1, z_2)$ and using the relations
\begin{gather}
    d(R z_1, R z_2) = R \frac{2 |z_1 - z_2|}{\sqrt{1 + R^2|z_1|^2}\sqrt{1 + R^2|z_2|^2}} = R \sqrt{h_R(z_1) h_R(z_2)} d(z_1,z_2), \qquad h_R(z) = \frac{1 + |z|^2}{1 + R^2 |z|^2} \\
    h_R(z_\xi(w)) = \frac{(1 + |\xi|^2)(1 + |w|^2)}{(1 + |\xi|^2)(1 + |w|^2) + (R^2 - 1)|w + \xi|^2} = \frac{1}{1 + \frac{R^2-1}{4} d^2(w,-\xi)} \\
\end{gather}
which together, yields
\begin{equation}
    d(R z_\xi(w_1), R z_\xi(w_2)) = R \frac{d(w_1, w_2)}{\sqrt{1 + \frac{R^2-1}{4} d^2(w_1,-\xi)} \sqrt{1 + \frac{R^2-1}{4} d^2(w_2,-\xi)}}.
\end{equation}
Substituting in (\ref{PotentialPairCorrelation}) or (\ref{PotentialSouthPole}) yields
\begin{align}
    v_\xi^R &= \int d\mu(z_1) d\mu(z_1) V\left(R \sqrt{h_R(z_1) h_R(z_2)} d(z_1, z_2)\right) g_\xi(z_1, z_2) \nonumber \\
    &= \int d\mu(w_1) d\mu(w_1) V\left(R \frac{d(w_1, w_2)}{\sqrt{1 + \frac{R^2-1}{4} d(w_1,-\xi)^2} \sqrt{1 + \frac{R^2-1}{4} d(w_2,-\xi)^2}}\right) g_{\rm SP}(w_1, w_2)
\end{align}
This expression allows us to compute the effective potential felt by quasiholes and the resulting dispersion, using Eq.~\ref{BersteinInversion}, for any interaction for finite or infinite system once we know $g_{\rm SP}(w_1, w_2)$. In the following, we will discuss further simplifications for specific interactions or in the thermodynamic limit.

\subsubsection{Coulomb interaction}

The above expression immediately implies the following quasihole energy for Coulomb interaction:
\begin{align}
    v_\xi^R &= \int d\mu(z_1) d\mu(z_1) \frac{1}{R d(z_1, z_2) \sqrt{h_R(z_1) h_R(z_2)}} g_\xi(z_1, z_2) \nonumber \\
    &= \int d\mu(w_1) d\mu(w_2) \frac{ \sqrt{1 + \frac{R^2 - 1}{4} d^2(w_1,-\xi)} \sqrt{1 + \frac{R^2 - 1}{4} d^2(w_2,-\xi)}}{R } g_{\rm SP}(w_1, w_2)
    \label{PotentialCoulomb}
\end{align}

\subsubsection{Pseudopotentials}
Psuedopotentials are expansions of short-range interactions in terms of derivatives of the delta function. For convenience, let us collect the results on the curved space that we will use. First, the integration measure acquires a metric dependence.
\begin{equation}
    d\mu(z) = \frac{4}{(1 + |z|^2)^2} dx dy
\end{equation}
In order for the delta function to satisfy the expected property
\begin{equation}
    \int d\mu(z) \delta_g(z,z_0) f(z) = f(z_0)
\end{equation}
its metric dependence should cancel the part in the integration measure
\begin{equation}
    \delta_g(z,z_0) = \frac{1}{\sqrt{g}} \delta(z - z_0) = \frac{(1 + |z|^2)^2}{4} \delta(z - z_0),
\end{equation}
where $\delta(z-z_0) = \delta(x - x_0) \delta(y-y_0)$.
This can be written in terms of $\gamma(z,z_0)$ defined in Eq.~\eqref{eq:stereographic-distance} by noting that
\begin{equation}
    \delta(\gamma(z,z_0)) = \frac{(1 + |z|^2)(1 + |z_0|^2)}{4} \delta(z - z_0) = \frac{(1 + |z|^2)^2}{4} \delta(z - z_0) = \delta_g(z,z_0)
\end{equation}
The Laplace operator~(Eq.~\eqref{eq:stereographic_laplacian}) takes the form
\begin{equation}
    \Delta_g =  (1 + |z|^2)^2 \partial_z \partial_{\bar z}.
\end{equation}
It satisfies
\begin{equation}
    \int d\mu(z) g(z) \Delta f(z) = \int d\mu(z) f(z) \Delta g(z)
\end{equation}
In particular, this implies the identity
\begin{equation}
    \int d\mu(z) f(z) \Delta^n \delta_g(z,z_0) = \Delta^n f(z) \Big|_{z = z_0}
\end{equation}
which implies
\begin{equation}
    \int d\mu(z) d(z,z_0)^{2m} \Delta^n \delta_g(z,z_0) = \begin{cases}
        (n!)^2 &: m = n \\
        0 &: m > n
    \end{cases}
\end{equation}

Under a dilation $z \mapsto R z$, the metric changes by $\sqrt{g(z)} \mapsto \sqrt{g(R z)} = R^2h_R(z)^2 \sqrt{g(z)}$. which means that 
\begin{equation}
    \delta_g(z,z_0) \mapsto \frac{1}{R^2 h_R(z)^2} \delta_g(z,z_0), \qquad d\mu(z) \mapsto R^2 h_R(z)^2 d\mu(z), \qquad \Delta \mapsto \frac{1}{R^2 h_R(z)^2} \Delta
\end{equation}
Thus, the energy for the pseudopotential $V_3(z_1,z_2) = \frac{1}{3!^2} \Delta^3 \delta_g(z_1,z_2)$ is given by
\begin{align}
    v_\xi^R &= \int d\mu(z_1) d\mu(z_2) V(R z_1, R z_2) g_\xi(z_1, z_2) \nonumber \\
    &= \frac{1}{3!^2} \int d\mu(z_1) d\mu(z_2) \frac{1}{R^8 h_R(z)^8} (\Delta^3 \delta_g(z_1,z_2)) d(z_1,z_2)^6 f_\xi(z_1,z_2) \nonumber \\
    &= \int d\mu(z) \frac{1}{R^8 h_R(z)^8} f_\xi(z,z) = \frac{1}{R^8} \int d\mu(w) \left(1 + \frac{R^2 - 1}{4} d^2(w,-\xi)\right)^8  f_{\rm SP}(w,w)
    \label{PotentialV3}
\end{align}

\section{Anyon dispersion in the thermodynamic limit}
\begin{figure}
    \centering
    \includegraphics[width=0.5\linewidth]{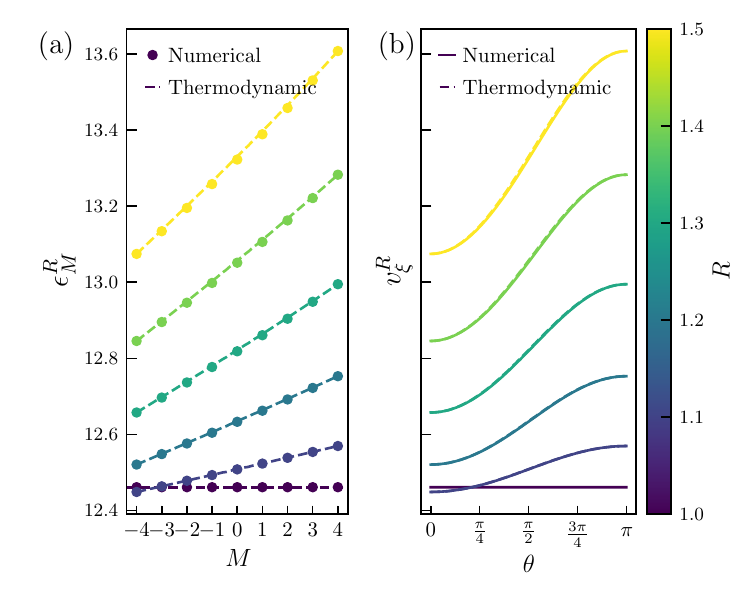}
    \caption{The comparison between anyon dispersion with Coulomb interaction and the analytic formula in Eq.~\eqref{eq:coulomb_analytic_dispersion}. The value of $\alpha(R)$ and $\beta$ were obtained from fitting ED result with the analytic formula.
    Remarkably, even though we are using Coulomb interaction, and therefore the ground states are not given by the exact zero modes of $V_1$ pseudopotential interaction, the analytic result agrees well with the numerical result.
    The ED was computed using $N_e = 8$ electrons. 
    }
    \label{fig:analytic_comparison_coulomb}
\end{figure}

A main simplification for evaluating the anyon dispersion in the thermodynamic limit is the following. As the system size $s$ increases (for a fixed radius of the sphere), the magnetic length $l_B \sim 1/\sqrt{s}$ decreases. The pair correlation function $g_\xi(z_1, z_2)$ is a function of the distances $d(z_1, z_2)$, $d(z_1, \xi)$ and $d(z_2, \xi)$. For each pair of variables, we expect the function to depend on this distance only up to few magnetic lengths and then to saturate to its value when such pair of variables are very far away. This follows from the plasma analogy where the charges are screened over a distance $\sim l_B$ \cite{}. Thus, we can write
\begin{equation}
    g_\xi(z_1, z_2) = F\left(\frac{d(z_1, \xi)}{l_B}, \frac{d(z_2, \xi)}{l_B}, \frac{d(z_1, z_2)}{l_B}  \right)
\end{equation}
Now clearly, if both $z_1$ and $z_2$ are far away from $\xi$, then the dependence on $\xi$ drops from this function and we get a constant contribution to $v_\xi^R$ which does not contribute to the dispersion. Thus, we only get contributions to the dispersion if either or both $z_1$ and $z_2$ are close to $\xi$. This contributions are expected to be negative compared to the asymptotic value of the function since they correspond to a dip where $z_{1,2}$ approaches the position of the quasihole. We distinguish two cases.
\begin{enumerate}
    \item Short-range interaction $V(d(z_1, z_2)) = W_\eta(d(z_1, z_2)/\eta)$ with $\eta \lesssim l_B$:\\
    In this case, both $z_1$ and $z_2$ will be close to $\xi$. For non-uniform field, we have $d(R z_1, R z_2) = R \sqrt{h_R(z_1) h_R(z_2)} d(z_1, z_2)$ but the function $h_R(z)$ changes slowly (on the scale of the radius of the sphere), so we can replace $z$ inside $h_R(z)$ by $\xi$ up to corrections which are small in $l_B$. Effectively, this means the simplification $V(d(R z_1, R z_2)) = V(R h_R(\xi) d(z_1, z_2))$ which amounts to the replacement
    \begin{equation}
        \eta \mapsto \eta_\xi^R = \frac{\eta}{R h_R(\xi)} = \frac{1 + R^2 |\xi|^2}{R (1 + |\xi|^2)} \eta 
    \end{equation}
    We can then expand the interaction in pseudopotentials 
    \begin{equation}
        V(d(R z_1, R z_2)) = \sum_{n=0} (\eta_\xi^R)^{2(2n+2)} c_n \hat V_{2n+1}(z_1, z_2), \qquad \hat V_{m}(z_1, z_2) = \frac{1}{m!} \Delta^{m} \delta(d(z_1, z_2)) 
    \end{equation}
    Note here that, usually, short-range interactions are defined with a normalization that also depends on $\eta$. For instance, $V_\eta(d(z_1,z_2)) = \frac{1}{\eta^2} e^{-\frac{d(z_1,z_2)^2}{2\eta^2}}$ which ensures $V_{\eta \rightarrow 0}(d(z_1,z_2)) \propto \delta(d(z_1,z_2))$. However, our transformation only affects the $\eta$ that enters in the functional form combined with $d(z_1,z_2)$ but not the normalization. This means that the dependence on $\eta_\xi^R$ in the power series has an extra power of 2 compared to the standard expression. For instance, for $V_3$, we get $(\eta_\xi^R)^8$ not $(\eta_\xi^R)^6$ which is also consistent with Eq.~\ref{PotentialV3}
    
    Each of the terms $\hat V_{2n+1}$ is rotationally symmetry and corresponds to the pseudopotential in the uniform field case. Since the function $g_\xi(z_1, z_2)$ is also rotationally symmetric, the integral $\int d\mu(z_1) d\mu(z_2) V_{2n+1}(z_1, z_2) g_\xi(z_1, z_2)$ is independent of $\xi$. Thus, all the energy dependence is encoded in the factors $\eta_\xi^R$, leading to the expression
    \begin{equation}
        v_\xi^R = a_0(R) - \sum_{n=1} (\eta_\xi^R)^{2(2n+2)} a_n c_n
    \end{equation}
    where $a_0(R)$ comes from the asymptotic value of $g_\xi(z_1,z_2)$ where they are far away from $\xi$. The minus sign encodes the fact that the deviation of $g$ from its asymptotic value is expected to have a negative integral. If our residual interaction is $\hat V_3$, such that $c_1 = 1$ and $c_{n>1} = 0$, this further simplifies to
    \begin{equation}
        v_\xi^R = a_0(R) - a_3 \left(\frac{1 + R^2 |\xi|^2}{R (1 + |\xi|^2)}\right)^8,  
    \end{equation}
    Thus, the spatial variations in the potential felt by the quasihole is determined fully analytically up to an overall multiplicative constant and a shift. Using Eq.~\ref{InversionThermodynamic} leads to the dispersion
    \begin{equation}
        \epsilon^R_{\kappa} = a_0(R) - a_3 \left(\frac{1+\kappa + R^2(1-\kappa)}{2R}\right)^8
    \end{equation}
    
    \item Coulomb interaction:\\ 
    For Coulomb interaction, the dominant contribution to the integral (\ref{PotentialCoulomb}) comes from having one of $z_{1,2}$ close to $\xi$ and one far away (having both close to $\xi$ have a smaller phase space). Without loss of generality, we can choose $z_1$ to be far away from $\xi$ and $z_2$ to be close to $\xi$. Then we can make the replacements
    \begin{equation}
        g_\xi(z_1, z_2) = F(\infty, d(z_2,\xi)/l_B, \infty), \qquad d(R z_1, R z_2) = R \sqrt{h_R(z_1) h_R(\xi)} d(z_1, \xi)
    \end{equation}
    The integral over $z_2$ can then be done giving a constant that does not depend on $\xi$ (since the integrand only depends on $d(z_2,\xi)$). Then we are left with the expression
    \begin{equation}
        v_\xi^R \propto \frac{1}{R \sqrt{h_R(\xi)}} \int d\mu(z) \frac{1}{\sqrt{h_R(z)} d(z,-\xi)} \propto \frac{\sqrt{1 + R^2|\xi|^2}}{R} \int d^2 z\frac{(1 + R^2 |z|^2)^{1/2}}{(1 + |z|^2)^{2}} \frac{1}{|z + \xi|}
    \end{equation}
    Noting that the integral depends only on $|\xi|$, we can assume $\xi$ is real and write
    \begin{align}
        I(R,\xi) &:= \int d^2 z\frac{(1 + R^2 |z|^2)^{1/2}}{(1 + |z|^2)^{2}} \frac{1}{|z + \xi|} = \int_0^\infty dr r \int_0^{2\pi} d\phi \frac{(1 + R^2 r^2)^{1/2}}{(1 + r^2)^{2}} \frac{1}{\sqrt{r^2 + \xi^2 + 2 r \xi \cos \phi}} \nonumber \\
        &= 4 \int_0^\infty dr r \frac{(1 + R^2 r^2)^{1/2}}{(1 + r^2)^{2} (r + \xi)} K \left(\frac{4 r \xi}{(r + \xi)^2} \right)
    \end{align}
    where $K(r)$ is the elliptic integral of the first kind. This yields the expression for the potential
    \begin{equation}
        v_\xi^R = \alpha(R) - \beta \frac{\sqrt{1 + R^2|\xi|^2}}{R} \int_0^\infty dr r \frac{(1 + R^2 r^2)^{1/2}}{(1 + r^2)^{2} (r + \xi)} K \left(\frac{4 r \xi}{(r + \xi)^2} \right)
        \label{eq:coulomb_analytic_dispersion}
    \end{equation}
    The dispersion is obtained by simply substituting $\xi = \sqrt{\frac{1-\kappa}{1+\kappa}}$ in this equation. The comparison of this dispersion with Coulomb dispersion is shown in Fig.~\ref{fig:analytic_comparison_coulomb}.
\end{enumerate}

\end{document}